\documentclass{article}

\usepackage[T1]{fontenc}
\usepackage[utf8]{inputenc}
\usepackage{microtype}
\usepackage{graphicx}
\usepackage{subcaption}
\usepackage{booktabs} 

\usepackage{hyperref}

\usepackage[preprint]{icml2026}

\usepackage{algorithm}
\usepackage{algorithmic}
\usepackage{float} 
\usepackage{mathtools}
\usepackage{amsthm}

\usepackage{amsmath}
\usepackage{amssymb}
\usepackage[dvipsnames]{xcolor}
\usepackage[most]{tcolorbox}
\usepackage{multirow}
\usepackage{colortbl}
\usepackage{tikz}
\usepackage{pifont}
\usepackage{textcomp}
\usepackage{fontawesome5} 
\usetikzlibrary{calc,positioning,arrows.meta,decorations.pathreplacing}

\usepackage{listings}

\definecolor{commentcolor}{RGB}{120,120,120} 
\definecolor{keywordcolor}{RGB}{200,30,30}   
\definecolor{tacticcolor}{RGB}{0,80,200}     
\definecolor{sortcolor}{RGB}{0,140,0}        
\definecolor{attributecolor}{RGB}{200,30,30} 
\definecolor{symbolcolor}{RGB}{0,0,0}        
\definecolor{numbercolor}{RGB}{140,140,140}  

\lstdefinestyle{lean}{
  language=lean,
  basicstyle=\ttfamily\scriptsize, 
  columns=fullflexible,
  keepspaces=true,
  showstringspaces=false,
  breaklines=true,
  breakatwhitespace=true,
  numbers=none,
  numberstyle=\ttfamily\tiny\color{numbercolor},
  frame=none,
  escapechar=§,
  xleftmargin=0.8em,
  keywordstyle=[1]{\ttfamily\color{keywordcolor}}, 
  keywordstyle=[2]{\ttfamily\color{sortcolor}},    
  keywordstyle=[3]{\ttfamily\color{tacticcolor}},  
  commentstyle=\itshape\color{commentcolor},
  identifierstyle=\ttfamily\color{black},
  stringstyle=\ttfamily\color{black},
}

\lstdefinestyle{python}{
  language=Python,
  basicstyle=\ttfamily\small,  
  columns=fullflexible,
  keepspaces=true,
  showstringspaces=false,
  breaklines=true,
  breakatwhitespace=true,
  numbers=none,
  numberstyle=\ttfamily\tiny\color{numbercolor},
  frame=none,
  xleftmargin=0.8em,
  keywordstyle=\ttfamily\color{keywordcolor}, 
  stringstyle=\color{tacticcolor},            
  commentstyle=\itshape\color{commentcolor},  
  emph={self},                                
  emphstyle=\color{sortcolor}
}

\definecolor{lightgreen}{HTML}{C8E6C9}
\definecolor{lightblue}{HTML}{E3F2FD}
\definecolor{darkgreen}{RGB}{0,115,0}

\usepackage[capitalize,noabbrev]{cleveref}

\theoremstyle{plain}

\theoremstyle{definition}

\theoremstyle{remark}

\usepackage[disable,textsize=tiny]{todonotes}

\icmltitlerunning{WybeCoder: Verified Imperative Code Generation}

\begin{document}
\twocolumn[
  \icmltitle{WybeCoder: Verified Imperative Code Generation}



\icmlsetsymbol{equal}{*}

  \begin{icmlauthorlist}

\icmlauthor{Fabian Gloeckle}{equal,meta,enpc}
\icmlauthor{Mantas Bakšys}{equal,meta,cambridge}
\icmlauthor{Darius Feher}{meta,ucl}
\icmlauthor{Kunhao Zheng}{meta,miles}
\icmlauthor{Amaury Hayat}{enpc,kias}
\icmlauthor{Sean B. Holden}{cambridge}
\icmlauthor{Gabriel Synnaeve}{meta}
\icmlauthor{Peter O'Hearn}{meta,ucl}

\icmlaffiliation{enpc}{CERMICS, ENPC, Institut Polytechnique de Paris, CNRS}
\icmlaffiliation{cambridge}{Computer Lab, University of Cambridge}
\icmlaffiliation{kias}{Korea Institute for Advanced Study}
\icmlaffiliation{meta}{FAIR, Meta}
\icmlaffiliation{miles}{Miles, LAMSADE, Université Paris-Dauphine-PSL, Paris}
\icmlaffiliation{ucl}{University College London}
  
  \end{icmlauthorlist}

  \icmlcorrespondingauthor{Fabian Gloeckle}{fgloeckle@meta.com}

  \icmlkeywords{Machine Learning, ICML}

  \vskip 0.3in
]



\printAffiliationsAndNotice{\icmlEqualContribution}

\begin{abstract}

Recent progress in large language models (LLMs) has advanced automatic code generation and formal theorem proving, yet software verification has not seen the same improvement. To address this gap, we propose WybeCoder, an agentic code verification framework that enables \textit{prove-as-you-generate} development where code, invariants, and proofs co-evolve.
It builds on a recent framework that combines automatic verification condition generation and SMT solvers with interactive proofs in Lean. 
To enable systematic evaluation, we translate two benchmarks for functional verification in Lean -- Verina and Clever -- to equivalent imperative code specifications. On complex algorithms such as Heapsort, we observe consistent performance improvements by scaling our approach, synthesizing dozens of valid invariants and dispatching dozens of subgoals, resulting in hundreds of lines of verified code, overcoming plateaus reported in previous works.
Our best system solves 74\% of Verina tasks and 62\% of Clever tasks at moderate compute budgets, significantly surpassing previous evaluations and paving a path to automated construction of large-scale datasets of verified imperative code.

\end{abstract}

\section{Introduction}

\begin{figure}[htbp]
    \centering
    \resizebox{\columnwidth}{!}{\input{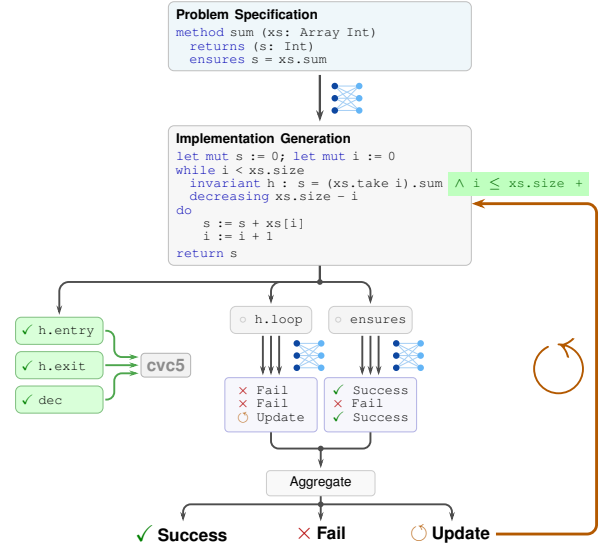}}
   \caption{WybeCoder: Subgoal decomposition multi-agent system. Starting from a problem specification, the agent generates an implementation and attempts to discharge verification conditions using CVC5. Remaining goals are tackled interactively in Lean, driving iterative implementation refinement or completing the proof.}
    \label{fig:decomposition}
\end{figure}

Large language models (LLMs) have made impressive progress in recent years in two related fields: code generation and interactive theorem proving. In both, they have advanced from toy problems in programming and mathematics to rivaling top human performance in prestigious competitions such as the IOI and the IMO \citep{alphaproof, alphacode}.

In software engineering, this progress has resulted in wide adoption of LLMs to speed-up code generation but still requires human review to ensure safety and quality standards. While scaling up code-generation is straightforward, reviewing remains time-consuming and labour intensive. This discrepancy exposes the need for better automated verification tools. Already commonly used approaches in verification include automated unit testing and fuzzing but provide only partial assurance by exposing defects (see Appendix~\ref{app:test-proof-gap}). Strong end-to-end guarantees require formal software verification: expressing intended behavior as a formal specification and constructing a machine-checked proof that the implementation satisfies it. For theorem proving, language models coupled with interactive theorem provers now surpass the performance of heuristic-guided automated theorem provers and SMT solvers \citep{alphaproof, achim2025aristotleimolevelautomatedtheorem}.

Existing works on LLMs in software verification have focused on two separate approaches: (a) verification of functional programs in interactive theorem provers, which leverages the impressive theorem proving capabilities of LLMs. However, this misses the reality of most existing software relying on imperative programming \citep{scott2025programming} that involves mutable state and side effects, and (b) verification of imperative code using “auto-active” environments such as Dafny \citep{leino2010dafny}, where agents generate ghost-code annotations for the program and SMT solvers handle the remainder of the proof, which matches the utility of imperative code but misses opportunities presented by LLMs as theorem provers.

In this work, we address this gap by developing a framework for verified imperative code generation in a \emph{hybrid} system where the agent recursively verifies and updates the generated code using a combination of SMT solvers and interactive prover agents in Lean~\citep{lean4}, enabled by recent software verification frameworks such as Loom \cite{gladshtein2026loom}. Our framework, WybeCoder, is named in homage to Edsger Wybe Dijkstra, whose pioneering work in verification emphasized program and proof generation being done together \citep{humble}, illustrated with imperative programs \citep{dijkstra1976discipline}. 

At small inference budgets, our work substantially outperforms prior approaches, improving on the best Verina result from 18\% \citep{ye2025verina} to 55\% with 32 refinement attempts. Our best results are state of the art, reaching 74\% on Verina and 62\% on Clever-Loom. Qualitatively, we observe that WybeCoder successfully implements and verifies in-place Selection Sort, Kadane's algorithm, and Heapsort, but makes only partial progress on recursive Quicksort. Overall, our work marks a significant step forward in agentic software synthesis, moving us closer to scalable generation and verification of imperative code.

Our \textbf{contributions} can be summarized as follows:

\begin{itemize}
    \item We investigate best-of-both-worlds \textbf{hybrid software verification frameworks} as an alternative to fully interactive verification (e.g. in Lean) and auto-active verification (e.g. in Dafny) \textbf{for LLM-based verification}, and introduce the first translated benchmarks tailored to such environments (Section~\ref{sec:benchmarks}).
    \item We develop and study multiple workflows for large-scale inference on such software verification tasks, exhibiting \textbf{subgoal decomposition and goal-directed modification} as key components (Sections~\ref{sec:method} and \ref{sec:results}).
    \item Our developed system continues to make progress on challenging problems such as the verification of Heapsort, using hundreds of LLM calls, routinely managing dozens of subgoals and multiple hundreds of lines of code (see example in Listing~\ref{lst:heapsort}). 
\end{itemize}

Our code is publicly available at {\faGithub \hspace{0.3mm}} \href{https://github.com/facebookresearch/wybecoder}{\mbox{facebookresearch/wybecoder}}.

\section{Background}

\subsection{Formal Software Verification}
Formal software verification is the task of proving correctness properties about code.
Depending on the type of code and the correctness properties sought, various calculi and frameworks for formal software verification exist.

\emph{Functional correctness} properties are typically easiest to state and prove in functional programming languages. This is because computations in functional programming languages are \emph{pure} mathematical functions uniquely defined by their input-output behavior without side effects to track in a proof. Correctness properties can be stated as mathematical theorems about these functions and can typically be proved with induction proofs over the inductive data types used in functional programming such as lists and trees. Interactive theorem provers such as 
Rocq \citep{the_coq_development_team_2024_14542673} and Lean are built upon such functional foundations and support function definitions in functional style, correctness theorems and native execution or code extraction.  For example, the fact that list reversion is self-inverse could be stated in Rocq as follows as a mathematical property of a pure total function \texttt{rev}:

\begin{lstlisting}[
    style=lean, 
    numbers=none, 
    basicstyle=\ttfamily\small, % Increases font size
    aboveskip=2pt,              % Removes the gap above
    belowskip=0pt               % Removes the gap below
]
Lemma rev_involutive : forall (A : Type) 
  (l : list A), rev (rev l) = l.
\end{lstlisting}

Functional programming often incurs performance overheads compared to imperative approaches when solving equivalent tasks. This disparity arises because imperative code typically leverages arrays, which provide constant-time access to arbitrary elements. In contrast, functional code relies on immutable data structures such as lists and trees, resulting in linear and logarithmic access times, respectively. An instance and further explanation of this phenomenon is referred to as the ``Log N Penalties in Functional Programming'' in a standard reference on the formal verification of functional algorithms \citep{vfa}.

For verifying imperative software, mutable state must be tracked explicitly in correctness proofs. This is performed by the \emph{Hoare logic} calculus \citep{hoare1969axiomatic}, where judgements are expressed as triples
\[
\{P\}\ C\ \{Q\}
\]
where $P$ is a pre-condition that holds before the command, $C$ the command to run and $Q$ a post-condition that holds after the command.
Inference rules explain how Hoare triples can be derived for composed constructs such as sequences of commands, \texttt{if} clauses or \texttt{while} loops. For example, if invariant $I$ is maintained by a \texttt{while} loop body, then after the loop the loop condition is false and the invariant still holds:
\[
\frac{
\{\,I \land B\,\}\ C\ \{\,I\,\}
}{
\{\,I\,\}\ \mathrm{while}\ B\ \mathrm{do}\ C\ \{\,I \land \lnot B\,\}
}
\]
\emph{Weakest pre-condition calculus} \citep{dijkstra1976discipline} derives, for every program construct except \texttt{while} loops, the weakest pre-condition under which a given postcondition is guaranteed to hold.
This reduces software verification to three main tasks: providing specifications \emph{(spec engineering)}, supplying sufficiently strong loop invariants \emph{(proof-hint synthesis)}, and proving that these invariants are preserved by the loop \emph{(proof engineering)}.
Many verification frameworks delegate the first two tasks to the user, while discharging the resulting (often routine) verification conditions with off-the-shelf SMT solvers such as Z3 \citep{demoura2008z3} or CVC5 \citep{barbosa2022cvc5}. This approach is exemplified by systems such as Dafny \citep{leino2010dafny} or the Frama-C WP plugin \citep{cuoq2012framac}.



While we also take termination measures into account \emph{(total correctness)}, extensions to Hoare logic that explicitly model heap memory state \citep[\emph{separation logic,}][]{reynolds2002separation} or shared memory in concurrent settings \citep{ohearn2007concurrency} are beyond the scope of this work.


\subsection{Hybrid Verification Frameworks}
\label{sec:hybrid}
A major drawback of SMT-based software verification is the opacity of automated provers. Failures are uninterpretable and provide little guidance on whether and how to refine the invariant annotations, resulting in a challenging proof debugging phase \citep{mugnier2025impact}.

To address this, we adopt a \emph{hybrid} software verification framework that bridges the gap between automated verification condition generation and interactive theorem proving. This approach allows users to revert to manual proofs in Lean or Rocq when automated solvers reach their limits. Specifically, we employ the Loom/Velvet system \citep{gladshtein2026loom} integrated within the Lean interactive theorem prover, which processes imperative syntax and invariant annotations into proof obligations. These obligations are then discharged via the CVC5 prover or, if necessary, through manual step-by-step verification. An example of a Velvet proof is given in Listing~\ref{lst:power4}.

\section{Evaluating Imperative Code Verification}
\label{sec:benchmarks}

Existing benchmarks for verified code generation target functional Lean, specifically Verina \citep{ye2025verina} and Clever \citep{thakur2025clever}. Both remain challenging for frontier models: on Verina, Claude Opus 4.5 achieves only 10\% proof success rate with 16 refinement attempts \citep{zhang2026claude}, o4-mini solving 22.2\% of proof tasks with 32 refinement attempts on Verina-advanced and all models evaluated on Clever solve under 9\% of the proof tasks. We translate both benchmarks to equivalent specifications in Velvet to evaluate imperative code synthesis and verification tasks. Due to the benchmark set-up and our translation methodology (see Appendix \ref{app:translation}), our evaluation results are directly comparable on Verina but not on Clever. We denote the translated version of the Clever benchmark Clever-Loom.

\begin{lstlisting}[caption={Velvet's hybrid approach: automated verification via \texttt{loom\_solve} and interactive proofs. GPT-5 on \texttt{verina\_basic\_43}.},label={lst:power4}, captionpos=t, language=lean, breaklines=true, style=lean]
method sumOfFourthPowerOfOddNumbers (n : Nat) return (result : Nat)
  ensures 15 * result = n * (2 * n + 1) * (7 + 24 * n^3 - 12 * n^2 - 14 * n)
  do
    -- Accumulate S_i = sum_{k=0}^{i-1} (2k+1)^4 using a loop
    let mut i : Nat := 0
    let mut acc : Nat := 0
    while i < n
      invariant h_i_le : i ≤ n
      -- Closed-form invariant for the partial sum (factored, avoiding powers)
      -- 15 * acc = i * (2*i - 1) * (2*i + 1) * (12*i*i - 7)
      invariant h_acc_closed : 15 * acc = i * (2 * i - 1) * (2 * i + 1) * (12 * i * i - 7)
      done_with h_done : n ≤ i
      decreasing h_dec : n - i
    do
      let odd : Nat := 2 * i + 1
      acc := acc + odd * odd * odd * odd
      i := i + 1
    return acc

lemma mul_factor_identity_nat (n : ℕ) :
  n * (2 * n - 1) * (2 * n + 1) * (12 * n * n - 7)
    = n * (2 * n + 1) * (7 + 24 * n ^ 3 - 12 * n ^ 2 - 14 * n) := by
sorry  -- 82 line proof omitted here

prove_correct sumOfFourthPowerOfOddNumbers by
  loom_solve
  case «ensures» => (
  cases i_3
  have hi_eq : i = n := Nat.le_antisymm h_i_le h_done
  have h1 :
      15 * acc
        = n * (2 * n - 1) * (2 * n + 1) * (12 * n * n - 7) := by
    simpa [hi_eq] using h_acc_closed
  exact h1.trans (mul_factor_identity_nat n)
  )
\end{lstlisting}

\subsection{Specification Leakage}
As observed by \citet{thakur2025clever}, especially for easier problems such as several of the considered benchmark problems, it can happen that the specification itself contains a computable function that determines the expected output. In such cases, the implementation can be retrieved from the specification and the proof reduces to reflexivity. In other words, \emph{specification leakage} renders the task trivial. An example for this behavior taken from one of our experiments is given below.
This issue is common in the Verina benchmark, which complicates the interpretation of evaluation results\footnote{\citet{harmonic2025verina} has reported 96.8\% performance on Verina. Without controlling for specification leakage and without being able to examine the results, it remains unclear to what extent such results reflect genuine advances in code synthesis}. We direct the reader to Section~\ref{sec:impjudge} for a discussion around mitigating specification leakage in our evaluations. 

\begin{lstlisting}[caption={Example of specification leakage on \texttt{verina\_basic\_57}},label={lst:spec-leak}, captionpos=t, language=lean, breaklines=true, style=lean]
method CountLessThan (numbers : Array Int) (threshold : Int) return (result : Nat)
  ensures result - numbers.foldl (fun count n => if n < threshold then count + 1 else count) 0 = 0 ∧
    numbers.foldl (fun count n => if n < threshold then count + 1 else count) 0 - result = 0
  do
    let res := numbers.foldl (fun count n => if n < threshold then count + 1 else count) 0
    return res
\end{lstlisting}

While the Clever benchmark explicitly acknowledges this limitation \citep{thakur2025clever} and attempts to mitigate it by using non-computable specifications, this mitigation is incomplete. For instance, the ``non-computable'' specification of Fibonacci numbers defines the graph of the function as an inductive relation, making it straightforward to recover the recursive function definition.

In the realm of imperative code verification, we therefore go another way for avoiding specification leakage. We allow functional leaking specifications in the dataset, and instead enforce imperative code in the implementations. For this, we need to solve the related issue of \emph{functional leakage} in the Velvet framework.

\paragraph{Functional leakage.}
\label{sec:func-leakage}
Due to Velvet's nature of being a \emph{shallow monadic embedding} in Lean, i.e. Velvet functions are terms in a special monad inside Lean, any pure Lean function can directly be used inside Velvet programs. This is part of the strength of the system as no separate standard library is required, but also allows cheating on imperative coding tasks by falling back to a fully functional solution wrapped inside the monad using \texttt{return} (the monad's \emph{unit}). The example from one of our experiments given below shows how such \emph{functional leakage} can happen even in the absence of specification leakage.

\begin{lstlisting}[caption={Example of functional leakage on \texttt{verina\_basic\_33}},label={lst:func-leak}, language=lean, breaklines=true, style=lean]
method smallestMissingNumber (s : List Nat) return (result : Nat)
  require List.Pairwise (· ≤ ·) s
  ensures ¬ List.elem result s ∧ (∀ k : Nat, k < result → List.elem k s)
  do
    let res := List.foldl (fun acc x => if x = acc then acc + 1 else acc) 0 s
    return res
\end{lstlisting}

\paragraph{Robustly mitigating leakage.}
We address both specification and functional leakage through a unified strategy: 
rather than attempting to obfuscate specifications as in Clever, we permit 
functional specifications but enforce imperative implementations. This targets 
a strictly harder task that is robust to both forms of leakage. Specifically, we 
convey detailed rules in the language model prompts and enforce them using 
LLM-based evaluation. We require solutions not to have runtime complexity 
contributions from functional primitives, i.e.\ to only have constant-time 
primitives on critical paths. \emph{Ghost variables} and invariants are permitted 
to use functional primitives, as their role is to couple mutable state with 
functional semantics rather than to contribute to the implementation. The full 
rules are included in the prompts in Appendix~\ref{app:prompts}.

\section{Method}
\label{sec:method}
We study multiple agentic systems that couple large language models with Lean's compiler feedback in the Velvet verification framework. We consider a sequential baseline that alternates between generation and execution, and a multi-agent system based on subgoal decomposition with proof reconstruction and goal-directed method modification.

\subsection{Sequential Agent}
\label{sec:linear}
The most basic option to couple language models with environment feedback is by means of a sequential agent that proposes code to run and receives Lean's error messages upon failure. The (unsurprising) pseudo-code for sequential agents is given in Algorithm~\ref{alg:linear}. We use sequential agents with compiler feedback as fundamental building blocks for multi-agent systems, where the individual agents fulfill tasks such as implementation generation or lemma proving.

\subsection{Subgoal Decomposition}
\label{sec:decomp}
In auto-active verification, failures are addressed by the rising sea \citep{Grothendieck:RecoltesEtSemailles} of additional helper lemmas and finer-grained invariants -- without precise feedback about what is missing. In interactive verification, by contrast, proof failure directly motivates invariant changes: insufficient hypotheses call for stronger invariants, while unprovable conclusions require weaker ones. We design our multi-agent system around this interactive workflow. Concretely, we extract proof obligations as independent theorems and dispatch them to parallel prover agents. If all subgoals succeed, the full correctness theorem is reconstructed. Unlike standalone lemma synthesis, this captures partial progress at the subgoal level. Sequential subagents are launched on each such subgoal, and if all goals are proved, the full correctness theorem is reconstructed from the subgoal proofs with an optional sequential agent that fixes potential issues in the proof reconstruction due to parsing idiosyncrasies. 

Modifying the method implementation is not an independent action but has to be \emph{conflict-driven}: informed by a concrete necessity where an agent considers their proof obligation unprovable. To enable proof transfer across method modifications, we use Lean's extensive metaprogramming utilities to introduce named invariants with a deterministic naming scheme for generated verification conditions (e.g., \texttt{h\_count.loop} for the loop-preservation obligation of an invariant named \texttt{h\_count}). When the method is updated, subgoal proofs from the previous attempt can be matched to new obligations by name, preserving partial progress. In this case, the method implementation is updated eagerly, the new subgoals are extracted but, crucially, we can attempt to transfer subgoal proofs from the previous attempt according to a naming scheme that links them to the invariants from which the goals were generated. The algorithm is summarized in Algorithm~\ref{alg:decomposition}.

\begin{algorithm}[t]
\caption{Subgoal decomposition multi-agent proof system}
\label{alg:decomposition}
\begin{algorithmic}[1]
\REQUIRE Language model agent $M$, initial method $\text{method}_0$, agent budget $N$
\ENSURE Reconstructed proof or \texttt{FAILURE}
\STATE $\text{method} \gets \text{method}_0$
\WHILE{agent budget $N$ not used up}
  \STATE $\text{goals} \gets \textsc{extract\_goals}(\text{method})$
  \STATE $\text{results} \gets \{\, \textsc{run\_provers\_concurrently}(M,g) \mid g \in \text{goals} \,\}$
  \IF{$\forall\, r \in \text{results},\ r \text{ is \texttt{SUCCESS}}$}
    \STATE \textbf{return} $\textsc{reconstruct\_proof}(M, \text{results})$
  \ELSIF{$\exists\, r \in \text{results},\ r \text{ is \texttt{CHANGE}}$}
    \STATE $\text{changes} \gets \{\, r \in \text{results} \mid r \text{ is \texttt{CHANGE}} \,\}$
    \STATE $\text{method} \gets \textsc{implement}(M, \text{changes})$
  \ELSE
    \STATE \textbf{return} \texttt{FAILURE}
  \ENDIF
\ENDWHILE
\STATE \textbf{return} \texttt{FAILURE}
\end{algorithmic}
\end{algorithm}

\section{Results}
\label{sec:results}
\begin{table}[t]
\centering
\small
\caption{\textbf{Results on the Verina benchmark.} Budget is represented as the turn limit $\times$ the number of agents launched per problem. For the Sequential Agent, $k$ agents in parallel corresponds to pass@$k$.}
\label{tab:verina}
\begin{tabular}{@{}lcc@{}}
\toprule
\textbf{Method} & \textbf{Turns $\times$ Agents} & \textbf{Solve Rate}\textsuperscript{\dag} \\
\midrule
\multicolumn{3}{@{}l}{\textit{Baseline}} \\
\quad DS Prover V2 7B & $64 \times 1$ & 20.0 \\
\midrule
\multicolumn{3}{@{}l}{\textit{Sequential Agent}} \\
  \quad GPT-5  & $32 \times 16$ & 64.6 \\
  \quad Gemini 3 Pro & $32 \times 16$ & 55.6 \\
  \quad Claude 4.5 Sonnet & $32 \times 16$ & 63.3 \\
  \quad Claude 4.5 Opus & $32 \times 16$ & \bf{74.1} \\

\quad GPTOSS-120B & $16 \times 4$ & 30.2 \\
\midrule
\multicolumn{3}{@{}l}{\textit{Subgoal Decomposition}} \\
\quad GPT-5 & $8 \times 128$ & 57.7 \\
\quad Gemini 3 Pro & $8 \times 128$ & 57.7 \\
\quad Claude 4.5 Sonnet & $8 \times 128$ & 51.9 \\
\quad Claude 4.5 Opus & $8 \times 128$ & 66.7 \\
\quad GPT-5 + Goedel Prover & $8 \times 512$ & 40.7 \\
\bottomrule
\end{tabular}

\vspace{2pt}
\parbox{\columnwidth}{\footnotesize\textsuperscript{\dag} Solve rate reports the percentage of successful proofs and disproofs. Disproofs are taken from the best sequential-agent run for all methods; see Section~\ref{sec:disproving}.}
\end{table}

\begin{table}[t]
\centering
\small
\caption{\textbf{Results on the CLEVER benchmark.} We evaluate imperative code verification using workflows described in Section~\ref{sec:method}. Correctness is measured as joint correctness of model-generated implementation and proof. Results are not immediately comparable between functional and Loom/Velvet versions (see Section~\ref{sec:benchmarks}). Budget is represented as the turn limit $\times$ the number of agents launched for each problem, except for the baseline COPRA. For the Sequential Agent, $k$ agents in parallel corresponds to pass@$k$.}
\label{tab:clever}
\begin{tabular}{@{}lcc@{}}
\toprule
\textbf{Method} & \textbf{Turns $\times$ Agents} & \textbf{Solve Rate}\textsuperscript{\ddag} \\
\midrule
\multicolumn{3}{@{}l}{\textit{Baseline}} \\
\quad COPRA (Claude 3.7) & 600 seconds & 8.7 \\
\midrule
\multicolumn{3}{@{}l}{\textit{Sequential Agent}} \\
  \quad GPT-5  & $32 \times 16$ & 53.8 \\
  \quad Gemini 3 Pro & $32 \times 16$ & 32.8 \\
  \quad Claude 4.5 Sonnet & $32 \times 16$ & 59.6 \\
  \quad Claude 4.5 Opus & $32 \times 16$ & \textbf{62.1} \\
\midrule
\multicolumn{3}{@{}l}{\textit{Subgoal Decomposition}} \\
\quad GPT-5  & $8 \times 128$ & 46.6 \\
\quad Gemini 3 Pro & $8 \times 128$ & 34.2 \\
\quad Claude 4.5 Sonnet & $8 \times 128$ & 41.0 \\
\quad Claude 4.5 Opus & $8 \times 128$ & 57.8 \\
\bottomrule
\end{tabular}

\vspace{2pt}
\parbox{\columnwidth}{\footnotesize\textsuperscript{\ddag} Solve rate reports the percentage of successful proofs. Disproving is not used on CLEVER.}
\end{table}

We evaluate frontier LLMs on software verification of imperative code using the Lean-based framework described in Section~\ref{sec:hybrid} using the inference methods described in Section~\ref{sec:method}, sequential agent and subgoal decomposition multi-agent.
Tables~\ref{tab:verina} and \ref{tab:clever} show our main results on end-to-end software implementation and verification success rates on the Verina and Clever-Loom benchmarks.

Our results show that \textbf{(i)} hybrid verification is a very effective ingredient for software verification agents, raising evaluation scores at small inference budgets from 15\% on Verina to 55\%, \textbf{(ii)} performance shows consistent scaling without plateaus across several orders of magnitude (Section~\ref{sec:scale}), \textbf{(iii)} our LLM-based imperativeness evaluation is effective at reducing specification leakage (Section~\ref{sec:impjudge}), and \textbf{(iv)} the system scales to verifying complex algorithms such as Heapsort (Section~\ref{sec:heapsort}) and its current limitations can be clearly outlined (Section~\ref{sec:quali}).

\subsection{Single-Agent vs Multi-Agent}
\begin{figure}[!t]
\centering
\begin{subfigure}[b]{0.48\textwidth}
  \centering
  \includegraphics[width=\textwidth]{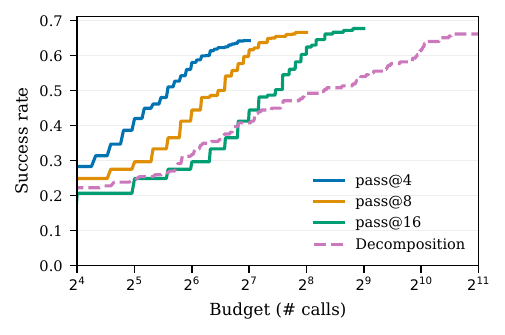}
  \caption{Claude 4.5 Opus}
  \label{fig:compare-opus}
\end{subfigure}
\hfill
\begin{subfigure}[b]{0.48\textwidth}
  \centering
  \includegraphics[width=\textwidth]{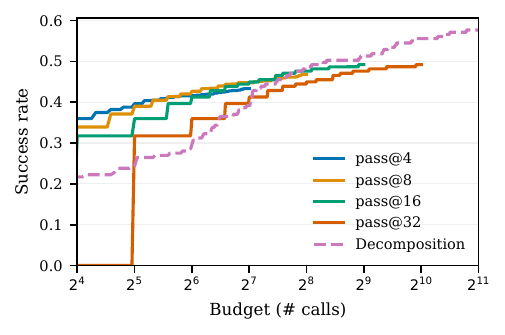}
  \caption{Gemini-3 Pro}
  \label{fig:compare-gemini}
\end{subfigure}
\caption{\textbf{Inference scaling for sequential agents and multi-agent system on Verina.} Different pass@k settings entail different compute-performance scalings for sequential agents which we compare to a multi-agent system. The optimal mechanism depends on the specific model in use, with Gemini-3 benefiting from the multi-agent scaffold while Claude 4.5 Opus does not.}
\label{fig:compare}
\end{figure}
Comparing the inference mechanisms of sequential agents and subgoal decomposition described in Section~\ref{sec:method}, we find:  performance is model-dependent with subgoal decomposition outperforming sequential agents by a significant margin for Gemini-3 while being surpassed for Claude 4.5 Opus, 4.5 Sonnet and GPT-5 (Figure~\ref{fig:compare}). 
Sequential agents are extremely efficient especially in the low-compute regime.
On larger-scale evaluations for sorting algorithms (Section~\ref{sec:quali}), we see anecdotal evidence for the scalability of the multi-agent system, continuously improving files over hundreds of LLM calls.
As an additional benefit, the progress of the multi-agent system is easily interpretable due to the breakdown by subgoals and \emph{conflict-driven} implementation updates. See Appendix~\ref{app:tradeoff} for a comparison to a naive multi-agent system.

\subsection{Sequential vs Parallel Compute}
\begin{figure}[!htbp]
  \centering
    \includegraphics[width=\linewidth]{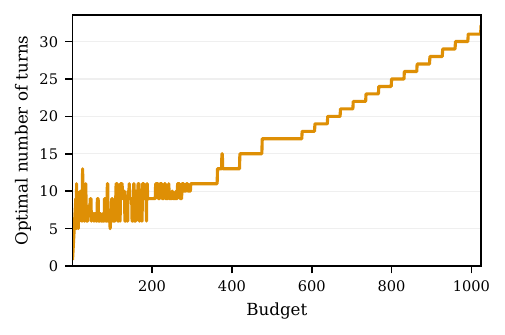}
    \label{fig:optimal-t-main}
  \caption{\textbf{Sequential vs Parallel Compute.} 
  Given a maximum inference budget of up to $C$ language model calls, we compute the optimal breakdown into $kT \leq C$ with $k \leq 32$ the number of independent attempts and $T \leq 32$ the maximum number of turns per attempt.
  Optimal $T$ ranges between 6 and 16 turns, increases beyond that are an artifact of bounded $k$ (Figure~\ref{fig:optimal}).}
  \label{fig:optimal-main}
\end{figure}

Sequential agents offer two directions to extend inference budgets: sequential compute (maximum number of turns~$T$) and parallel compute (number of independent attempts~$k$ in pass@$k$). As can be seen in Figure~\ref{fig:optimal-main}~and~\ref{fig:optimal}, there is a tradeoff between these two depending on the total available budget, with optimal allocation potentially influenced by language model post-training specifics.


\subsection{Disproving}
\label{sec:disproving}
Software verification benchmarks can have ``label errors'' in the sense of unsatisfiable specifications. Due to our translation methodology (see Appendix \ref{app:translation}), we made the decision for Clever-Loom to update the benchmark and fix such errors while for Verina, we aim for exact comparability and instead allow the agent to prove that the specification is unsatisfiable. For this, we automatically generate theorems of the form $\exists x \in X, P(x) \wedge \forall y \in Y, \neg Q(x, y)$ if $P$ is the pre-condition on inputs $x \in X$ and $Q(x, y)$ is the post-condition relating inputs $x$ with outputs $y \in Y$. The model is not informed whether the specification is satisfiable but has to decide whether prove or disprove it.

In 32-turn evaluations, all models except Claude 4.5 Sonnet proved 12 specifications to be unsatisfiable (Sonnet: 11), with GPT-5 notably successful on 347/384 of these cases. Note that unsatisfiability (regardless of implementation) is a stronger result than the 23 ground truth implementations that \citet{harmonic2025verina} prove incorrect.


\subsection{Imperativeness Judge and Functional Leakage}
\label{sec:impjudge}
In initial experiments, we observed a high degree of \emph{functional leakage}: formally imperative methods that are only a thin wrapper around an otherwise functional computation (Section~\ref{sec:func-leakage}).
Indeed, an evaluation of the subgoal decomposition agent on Verina that did not guard against this behavior resulted in a success rate of 75.1\% (GPT-5, 8-turn pass@64) and prompted us to include an LLM-based evaluation of \emph{imperativeness}, after which the performance dropped to 51.9\%, but with properly imperative programs according to our manual review. We use an imperativeness throughout in all reported scores, and note that as specifications are functional, this also prevents specification leakage.

\subsection{Scaling and Model Comparison}
\label{sec:scale}
\begin{figure}[!ht]
    \centering
    \includegraphics[width=\linewidth]{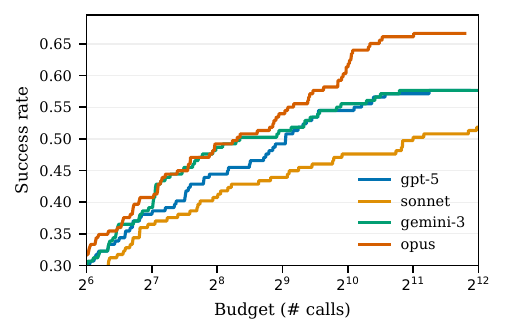}
    \caption{\textbf{Inference scaling comparison for multi-agent system with different models.}
    We evaluate using up to 128 subagents on Verina and plot by the total number of LLM calls spent according to iterative deepening (Section~\ref{app:tradeoff}). See Figure~\ref{fig:decomp-scale} for comparisons using latency and the Clever-Loom benchmark.
    }
    \label{fig:verina-decomp}
\end{figure}

Comparing the frontier models GPT-5, Gemini 3 Pro, Claude 4.5 Opus and Claude 4.5 Sonnet as a smaller model on both sequential agent and multi-agent inference, we find that the larger models are roughly matched with Opus best overall, all outperforming the smaller Sonnet model (Tables~\ref{tab:verina}~and~\ref{tab:clever}, Figure~\ref{fig:verina-decomp}). Across all models, there is no sign of plateauing performance across several order of magnitude of inference compute. For all experiments, the sampling temperature was set to 1.0, except for GPT-5, where the API did not expose temperature control.

\subsection{ Scalability of Multi-Agent Systems}
\begin{figure}[t]
    \centering
    \includegraphics[width=\linewidth]{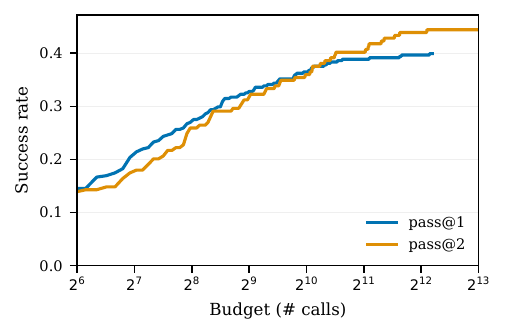}
    \caption{\textbf{Multiple attempts with multi-agent systems.}}
    \label{fig:decomp_at2}
\end{figure}
Scaling inference via multi-agent systems can outperform naive pass@k because agents can share state, so extra compute can be used to improve a single rollout rather than spawning independent replicas. To test how much this holds for our subgoal-decomposition system (Section~\ref{sec:decomp}), we ran GPT-5 on Clever with $k=2$ copies. Figure~\ref{fig:decomp_at2} shows that allocating additional compute to a single copy continues to improve performance up to 
~1200 model calls, after which pass@2 becomes preferable; in contrast, for sequential agents the crossover occurs much earlier, at ~22 calls.







\subsection{Qualitative Evaluation}
\label{sec:quali}
\begin{table}[h!]
\centering
\caption{\textbf{Sorting algorithms and their sub-procedures.} Ticks denote successful verification.}
\begin{tabular}{ll}
\toprule

Selection Sort & \checkmark \\
Bubble Sort & \checkmark \\
Insertion Sort & \checkmark \\
Binary Insertion Sort & \checkmark \\
Recursive Quicksort & $\times$ \\
Quicksort & \\
\hspace{2em}Partition & \checkmark \\
\hspace{2em}Step & $\times$ \\
\hspace{2em}Sort & \checkmark \\
Heapsort & \\
\hspace{2em}Heapify & \checkmark \\
\hspace{2em}Maxheap & \checkmark \\
\hspace{2em}Sort & \checkmark \\
Quicksort & \\
\hspace{2em}Mergeruns & \checkmark \\
\hspace{2em}Mergepass & $\times$ \\
\hspace{2em}Sort & \checkmark \\
\bottomrule
\end{tabular}
\label{tab:sorts}
\end{table}
By manually reviewing rollout successes and failures on Verina and Clever, we identify the limits of our system's verification capabilities. Employing moderate inference budgets frontier models consistently succeed at verifying constant-time algorithms (e.g. computing the area of a triangle), list scan- and fold-like operations (e.g. maximum, find) and naive brute-force algorithms (e.g. find pair). Tasks that require heavy use of array mutation or dynamic programming are comparatively rarely solved, or solved by employing Lean's libraries to reduce implementation and verification complexity. An example of this we have observed is the agent updating array elements using \texttt{Array.push} instead of by \emph{moving} elements as mandated by the description. This analysis puts sorting algorithms at the very edge of what is feasible with moderate compute budget with frontier models. When prompted to implement and prove a specific sorting algorithm, models frequently succeed in the cases of Insertion Sort, Selection Sort, Bubble Sort or Binary Insertion Sort, but fail on the more challenging examples of in-place imperative Quicksort, imperative Mergesort with $\mathcal{O}(n)$ additional memory or Heapsort. After breaking down the last three sorting algorithms manually into three sub-routines with specifications each (but without implementations), Heapsort completes successfully, while Quicksort and Mergesort each complete 2/3 sub-routines, leaving the most challenging one unverified. See Appendix~\ref{app:quali} for details.

\subsection{Full Verification of Heapsort}
\label{sec:heapsort}
The successful verification of Heapsort using 215 sub-agents for \texttt{heapify}, 8 for \texttt{maxheap} and 134 for \texttt{heapsort\_main} using our subgoal decomposition system and Claude 4.5 Opus demonstrates the scalability of our approach and the capabilities of frontier models, but also a certain degree of inefficiency. Heapsort is often considered a challenging example for verification. The program proving text of \citet{reynolds1981craft} gives a by-hand in a section which is starred to connote difficulty, and later formal treatments would sometimes take up an entire paper (e.g.,~\citet{tafat2011binary}). For the full code and more examples, see Appendix~\ref{app:examples}.

\section{Related Work}

\paragraph{LLMs for Auto-Active Verification}
Pioneering works have established the use of LLMs in software verification, initially focusing on single-turn proof-hint synthesis. Benchmarks such as DafnyBench \citep{loughridge2024dafnybenchbenchmarkformalsoftware} and VerifyThisBench \citep{deng2025verifythisbenchgeneratingcodespecifications} (Dafny/Why3), Verus-Bench \citep{chen2025automatedproofgenerationrust} and Verified Cogen \citep{verified-cogen} (Verus), and InvBench \citep{wei2025invbenchllmsaccelerateprogram} and LIG-MM \citep{liu2024towardsgeneralloop} (C/Java) have been used to evaluate the utility of LLMs in generating invariants and reachability proofs.

Building on these benchmarks, frameworks like Clover \citep{sun2024cloverclosedloopverifiablecode}, DafnyGym \citep{mugnier2025laurelunblockingautomatedverification}, and AutoVerus \citep{AutoVerus} use iterative feedback loops, demonstrating how compiler error messages can guide LLMs to refine proofs over multiple turns. While these studies primarily address the challenge of invariant and other proof-hint infilling for existing code, our work extends this scope to the end-to-end synthesis of implementation and verification in tandem. Additionally, while this work targeted introductory programming problems (such as the MBPP dataset), we benchmark our framework on implementation and verification of more complex algorithms.

\paragraph{Interactive Software Verification}
While LLMs are increasingly applied to software verification in interactive theorem provers, benchmarks like Verina \citep{ye2025verina}, Clever \citep{thakur2025clever}, and VeriBench \citep{anonymous2025veribench} focus primarily on functional Lean code.
Current frontier models exhibit a stark performance gap in these settings: while OpenAI’s o4-mini achieves 61.4\% on code generation (Verina pass@64), its proof synthesis task pass rate on Verina-advanced remains modest at 22.2\%.
Nevertheless, landmark verification success stories like the verification of the seL4 microkernel \citep{sel4} and CompCert \citep{compcert} demonstrate the potential and scalability of software verification using ITPs.
Furthermore, recent breakthroughs in IMO-level automated reasoning \citep{chen2025seedprover, achim2025aristotleimolevelautomatedtheorem} suggest that LLMs, when paired with robust inference scaling, can navigate complex formal proofs. Diverging from the functional focus of existing work, we introduce a novel approach targeting the \emph{interactive} verification of \emph{imperative} code with LLMs.

\section{Conclusion}
Following recent breakthroughs in formal mathematical theorem proving and verification of functional code with language models, fully-automated verification of imperative programs using Hoare logic is the next frontier in LLM-based software verification and will pave the way to real-world applications where memory layout (separation logic) and concurrency play a role too. 

In this study, we adopted a hybrid verification framework (Loom/Velvet) that combines the strengths of auto-active verification frameworks such as Dafny with interactive theorem proving in Lean. We demonstrated that current frontier language models can leverage such frameworks purely in-context, achieving proficiency comparable to their capabilities in functional code verification despite the added complexity of mutable state and loop invariants.

We designed, evaluated and compared different multi-agent inference workflows for effectively scaling inference compute at reasonable end-to-end verification latencies, and highlighted subgoal decomposition and goal-directed modification as ingredients of a particularly well-suited workflow for this task. Our best system solves 74\% of Verina and 62\% of Clever tasks, and successfully verifies algorithms requiring deep insight such as array-based Heapsort, raising the bar beyond mere computations and brute-force solutions.

With complexity-aware verification as a natural next step, our results suggest that large-scale automated construction of verified imperative code datasets is now within reach. This opens new avenues for training and reinforcement learning on provably correct programs, and brings us closer to a future where AI-generated code comes with formally-verified correctness guarantees.

\section*{Limitations}
While our results show that reasoning-focused LLMs bring automatically verified imperative code closer to reality, several limitations currently prevent production deployment.
First, our Loom/Velvet pipeline relies on an experimental, actively developing software package with custom modifications (Section~\ref{sec:hybrid}), and results may change as the framework matures.
Second, the generated imperative code targets Lean, which uses managed memory. Our approach does not yet address unmanaged production languages such as C++ or Rust, nor their memory models (e.g.\ separation logic, ownership types).
Third, our scope covers code generation and correctness proofs but omits the equally critical upstream tasks of specification engineering and sub-method decomposition for our sorting algorithm evaluation, which remain manual in our pipeline.
Finally, most of our experiments use closed-weight frontier models via API, with open-source models performing substantially worse potentially limiting reproducibility.

\section*{Impact Statement}

This paper presents work whose goal is to advance the field of Machine
Learning. There are many potential societal consequences of our work, none
which we feel must be specifically highlighted here.

\bibliography{references}
\bibliographystyle{icml2026}

\newpage
\appendix
\onecolumn
\section*{Appendix}

\section{Test and Proof Pass-Rate Gap}\label{app:test-proof-gap}

\begin{figure}[htbp]
  \centering
  \begin{subfigure}[b]{0.45\textwidth}
    \centering
    \includegraphics[width=\linewidth]{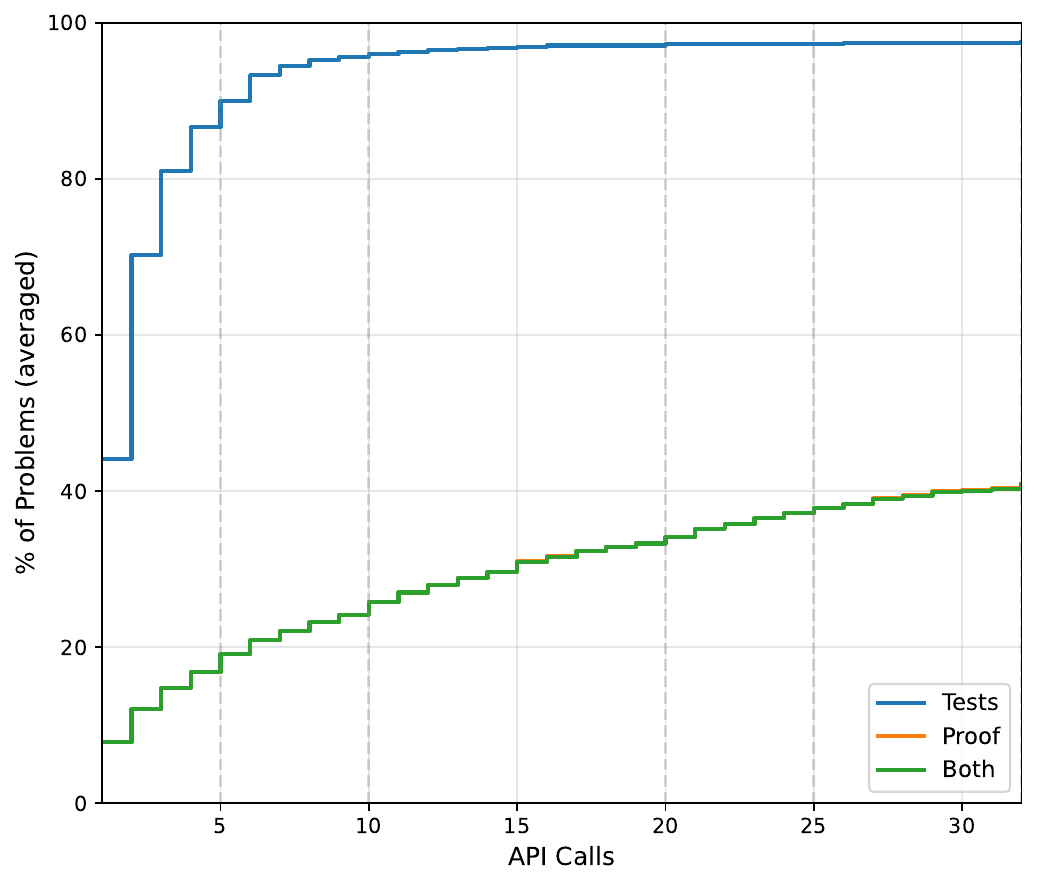}
    \caption{Clever}
    \label{fig:first}
  \end{subfigure}\hfill
  \begin{subfigure}[b]{0.45\textwidth}
    \centering
    \includegraphics[width=\linewidth]{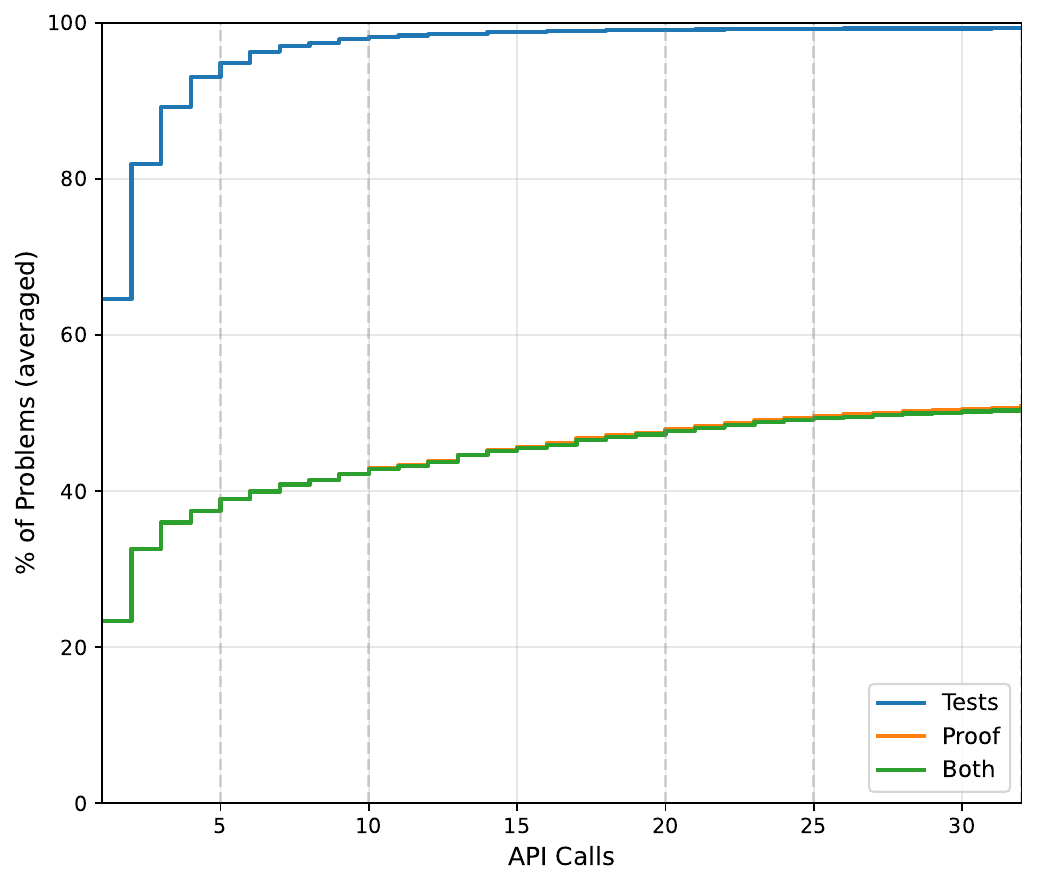}
    \caption{Verina}
    \label{fig:second}
  \end{subfigure}
  \caption{Per-problem cumulative pass rates vs. number of LLM API calls for (a) Clever and (b) Verina. Curves represent the fraction of problems achieving first success on tests, proofs, or both. Statistics are computed per-problem then averaged to avoid overweighting easier problems.}
  \label{fig:test_vs_proof_difficulty}
\end{figure}

Figure~\ref{fig:test_vs_proof_difficulty} shows that test success rises rapidly and approaches 100\%, whereas proof success improves much more slowly. The large and consistent gap between the test and proof curves suggests that proving correctness is substantially harder than passing tests.

\section{Benchmark Translation Methology}
\label{app:translation}

\subsection{Verina-Loom}
Verina provides structured metadata for each task, including separated function signatures, pre-conditions, and post-conditions. We automatically extract these components and assemble them into Velvet method specifications using Dafny-inspired syntax. This mechanical translation ensures strict comparability with existing results on functional code verification.
\subsection{Clever-Loom}
Clever specifications are monolithic propositions relating inputs and outputs without separating pre- and post-conditions. While these could in principle be used as monolithic post-conditions, doing so would undermine the efficiency of SMT-based verification, which excels at discharging many small obligations but struggles with large compound statements.
Rather than parsing the specifications to extract logical structure, we adopt a semi-automated pipeline:
\begin{enumerate}
    \item Prompt Gemini 2.5 Pro to translate each specification into Velvet \texttt{require}/\texttt{ensures} syntax
    \item Conduct systematic human review to ensure correctness and idiomatic style
    \item Fix unsatisfiable specifications from the original dataset where identified
\end{enumerate}
We aim to stay as close as possible to the original Clever specifications, but some deviations are necessary for idiomatic Velvet. As a result, benchmark numbers are not strictly comparable to the original Clever results. However, this process yields a high-quality benchmark for verified imperative code, with several specification errors from the original dataset corrected.

From Verina's structured data format, we retrieve the signature, pre-conditions and post-conditions, allowing us to assemble the corresponding Velvet method specification automatically. See Listing~\ref{lst:power4} for an example including the specification header in Dafny-inspired syntax. With the resulting Verina-Loom benchmark, we aim for strict comparability with existing numbers functional code verification.

\section{Tool Use}\label{app:tool-use}
In initial experiments, we identified missing knowledge about Lean's standard library (generally and in the specific version that Loom/Velvet uses) a limiting factor for LLM agents to produce correctness proofs. We therefore provide library search as \emph{tools} \citep{Schick2023Toolformer} for language models with functional calling capability. Specifically, we deploy Loogle \citep{BreitnerLoogle} and LeanExplore \citep{AsherLeanExplore2025} as local services, two tools for semantic and syntactic search in Lean's standard library and \texttt{mathlib} \citep{mathlib}.

Instead of a flat design where the agent decides at each step whether to run Lean code or perform a library search and all interactions are concatenated, we opt for a nested design where outer turns are interaction with Lean and inner turns are library search tools. The agent can perform multiple library search tool calls until the inner loop budget runs out or it decides to return to the outer Lean loop. The agent is requested to summarize its library search findings for subsequent rounds, and the inner loop messages are removed from the chat history in order to save context tokens and refocus on the main verification task.

Empirically, tool-augmented library search does not improve pass rates over the baseline, but it increases compute because the agent spends additional turns querying tools and summarising results before returning to the main Lean loop.

\section{Extended Related Work}
\label{app:related-work}


\subsection{Agentic software verification}

\paragraph{Auto-Active Verification: Datasets and Evaluation}
In auto-active settings, verification relies on SMT solvers (e.g., CVC5 or Z3) to discharge proof obligations, shifting the agent's burden from full proof construction to \textit{proof hint inference}--specifically the synthesis of loop invariants and pre/post-conditions.
\textsc{DafnyBench} \citep{loughridge2024dafnybenchbenchmarkformalsoftware} provides the largest corpus for this task, comprised of verified Dafny programs stripped of proof hints where agents must supply missing invariants and assertions to verify existing code. Similarly, \textsc{VerifyThisBench} \citep{deng2025verifythisbenchgeneratingcodespecifications} aggregates tasks across Dafny and Why3, focusing on the joint synthesis of specifications and implementations.
More recent efforts target system-level languages, \textsc{Verus-Bench} \citep{yang2025verusagestudyagentbasedverification} and \textsc{Verified Cogen}\citep{verified-cogen} evaluate the synthesis of verified Rust code via the Verus toolchain, testing the model's ability to navigate complex memory safety constraints. These modern datasets build upon foundational competitions like SV-COMP \citep{sv-comp}, which continue to serve as rich sources for mining safety properties and program lemmas in C and Java. To this end, \textsc{InvBench} \citep{wei2025invbenchllmsaccelerateprogram} draws verification tasks from SV-COMP's reachability benchmarks to evaluate LLM-based invariant synthesis, while \citep{liu2024towardsgeneralloop} incorporate SV-COMP programs into their benchmark for loop invariant generation with memory manipulation in auto-active verification settings. Throughout this line of work, frontier LLMs have shown some emergent capability of aiding with tasks such as invariant, and implementation synthesis but the relative brittleness introduced by the SMT-backends has slowed progress of expanding scope and complexity of the verification tasks studied.

In addition to these datasets, recent systems increasingly treat the verifier as an interactive environment and use its feedback to iteratively repair annotations, specifications, and proofs. Clover \citep{sun2024cloverclosedloopverifiablecode} proposes a framework in Dafny to synthesize programs and annotations using natural language descriptions and applies it to relatively introductory MBPP-level problems with ground-truth implementations. Laurel \citep{mugnier2025laurelunblockingautomatedverification} further isolates this approach by focusing on assertion generation rather than end-to-end verified program synthesis. In parallel, Verus-based \citep{verus} agentic work starts from existing Rust code and generates specifications and proofs without a held-out ground-truth specification \citep{AutoVerus,chen2025automatedproofgenerationrust}, making its evaluation regime closer to spec-and-proof recovery than to end-to-end generation under a reference spec. In contrast, WybeCoder is an end-to-end verified imperative code generation in a hybrid environment, where interactive goals can drive invariant and proof refinement and where inference can be scaled via subgoal decomposition.

\paragraph{ITP-Based Verification: Datasets and Evaluation}
Conversely, interactive theorem proving requires the explicit construction of proof terms or tactic scripts, presenting a steeper challenge for current models. \textsc{Clever} \citep{thakur2025clever} provides a curated benchmark of 161 problems for end-to-end verified code generation in Lean 4, where each problem requires both generating a specification matching a held-out ground truth and producing a provably correct implementation. Evaluations reveal that state-of-the-art models achieve an end-to-end success rate of less than 1\%. \textsc{Verina} \citep{ye2025verina} offers a complementary benchmark of 189 programming challenges with modular evaluation across code, specification, and proof generation tasks. On this benchmark, the best-performing model (OpenAI o4-mini) achieves 61.4\% on code generation, 51.0\% on specification generation, but only 22.2\% on proof generation at pass@64. Beyond the code-generation focus of these benchmarks, \textsc{VeriBench} \citep{anonymous2025veribench} tests translation capabilities, requiring agents to convert Python algorithms into verified Lean 4 implementations including unit tests, correctness theorems, and formal proofs. In their evaluations, frontier LLMs' compilation rates of only 12.5\% contrast sharply with mathematical theorem proving performance, a substantial gap in neural methods for software verification applications in ITPs. Weak performance of current agentic systems in software verification is not an inherent limitation of the ITP framework: landmark projects such as the seL4 microkernel verification in Isabelle \citep{sel4}, CompCert in Rocq \citep{compcert}, and AWS Nitro Isolation engine verification \citep{aws_graviton5_2025} with Isabelle/HOL and Idris demonstrate that large-scale verified software is achievable with sufficient human expertise, promising a clear pathway for scaling multi-agent approaches for this task.

\subsection{Neural Formal Theorem Proving} 
The application of LLMs to formal mathematics has evolved from simple tactic generation to sophisticated neural theorem-proving systems integrated with ITPs. Early efforts, beginning with GPT-f, demonstrated that LLMs could be trained to generate valid proof steps in Lean by training on a curriculum of formal statements and proofs \citep{polu2022formalmathematicsstatementcurriculum, polu2020generativelanguagemodelingautomated}. However, progress was initially constrained by the scarcity of high-quality formal data and the challenges of long-horizon reasoning.

To navigate the vast state-space in the formal theorem proving environment, early neural theorem provers primarily focused on exclusively formal code generation and heavily relied on external search algorithms like Monte-Carlo Tree Search \citep{lample2022hypertreeproofsearchneural}. However, recent breakthroughs in reasoning \citep{openai2024openaio1card} have enabled a tighter integration of informal and formal domains at inference time. By interleaving informal \textit{thinking traces} with formal code blocks, systems like DeepSeek-Prover-v2 \citep{ren2025deepseekproverv2advancingformalmathematical} and Kimina-Prover \citep{wang2025kiminaproverpreviewlargeformal} leverage the vast mathematical knowledge accumulated during pre-training while grounding their logic in verifiable rewards.

These advances have substantially improved long-horizon, tool-integrated formal reasoning, leading to large gains on established formal-math benchmarks \citep{chen2025seedprover, alphaproof}. Strong formal reasoners can also enable autoformalization and the construction of large verified corpora\citep{urban2026130klinesformaltopology, strongpnt2025}, helping alleviate data scarcity for training and evaluation. While primarily developed for formal mathematics, these tool-integrated reasoning and repair techniques are likely transfer to naturally to verified programming, where agents must additionally synthesize and refine proof-hints and loop invariants for 
programs. 

\section{Algorithms}
\label{app:algorithms}
We provide pseudocode for the inference workflows described in Section~\ref{sec:method}.\\
\\
\textbf{Sequential Agent (Algorithm~\ref{alg:linear}).} The workflow alternates between language model generation and Lean compiler feedback. The agent accumulates a history of (code, error) pairs and conditions each subsequent generation on this context until verification succeeds or the budget is exhausted.

\textbf{Lemma-Writing Multi-Agent System (Algorithm~\ref{alg:lemmas}).} Multiple agents work concurrently on a shared file, each proposing one of three actions: modify the implementation, conjecture and prove helper lemmas, or edit the final proof. Successful contributions are merged into the shared state for subsequently launched agents. We explored this system in early experiments but observed sub-par results compared to sequential and subgoal decomposition agents. We therefore did not scale this approach further.

\noindent
\begin{minipage}[t]{0.48\textwidth}
\begin{algorithm}[H]
\caption{Sequential agent}
\label{alg:linear}
\footnotesize
\begin{algorithmic}[1]
\REQUIRE Language model $M$, Lean environment $E$, problem specification $\mathcal{P}$, budget $T$
\ENSURE Verified Lean code or \texttt{FAILURE}
\STATE $\text{history} \gets [\,]$
\FOR{$t \gets 1$ \textbf{to} $T$}
    \STATE $\text{code} \gets M(\mathcal{P}, \text{history})$
    \STATE $\text{result} \gets E(\text{code})$
    \IF{$\text{result is \texttt{SUCCESS}}$}
        \STATE \textbf{return} $\text{code}$
    \ENDIF
    \STATE $\text{history} \gets \text{history} + [(\text{code}, \text{result.error})]$
\ENDFOR
\STATE \textbf{return} \texttt{FAILURE}
\end{algorithmic}
\end{algorithm}
\end{minipage}
\hfill
\begin{minipage}[t]{0.48\textwidth}
\begin{algorithm}[H]
\caption{Lemma-writing multi-agent proof system}
\label{alg:lemmas}
\footnotesize
\begin{algorithmic}[1]
\REQUIRE Language model agent $M$, Lean environment $E$, method specification $\mathcal{P}$, degree of parallelism $k$, agent budget $N$, initial method $\text{method}_0$, initial error $\text{error}_0$
\ENSURE Verified Lean code or \texttt{FAILURE}
\STATE $\text{file} \gets \textsc{make\_file}(\text{method}_0,\ \text{default\_proof})$
\STATE $\text{error} \gets \text{error}_0$
\STATE $k$-\textbf{parallel} \algorithmicfor\ $i \gets 1$ \textbf{to} $N$ \algorithmicdo
\begin{ALC@for}
  \STATE $\text{action} \gets M(E, \mathcal{P}, \text{file}, \text{error})$
  \STATE $\text{file} \gets \textsc{update\_file}(\text{file}, \text{action})$
  \STATE $\text{result} \gets E(\text{file})$
  \IF{$\text{result is \texttt{SUCCESS}}$}
    \STATE \textbf{return} $\text{file}$
  \ENDIF
  \STATE $\text{error} \gets \text{result.error}$
\end{ALC@for}
\STATE \algorithmicendfor
\STATE \textbf{return} \texttt{FAILURE}
\end{algorithmic}
\end{algorithm}
\end{minipage}

\section{Additional Quantitative Evaluation Results}

\subsection{Evaluating Multi-Agent Systems: Accuracy/Latency/Compute Tradeoffs}
\label{app:tradeoff}

Evaluations of machine learning systems should not merely report single performance numbers, but their scaling with respect to computational inputs. For single-agent LLM systems, the computational budget is approximated by the number of LLM calls (or tokens) used. For multi-agent setups, the presence of concurrency bifurcates this metric into two separate notions: \emph{compute}, i.e. the total amount of computational resources (calls, tokens) used overall, and \emph{latency}, i.e. the total amount of computational resources spent on the longest critical path, measured as a time duration or likewise in tokens or LLM calls.

Running a single-agent system $k$ times independently in parallel is a basic multi-agent system, namely pass@k evaluation. In this case, latency is the maximum of the lengths of all $k$ trajectories while compute is their sum.

Indeed, for conducting up to $k$ evaluations with early stopping, different parallelism schemes allow to obtain different latency-compute tradeoffs: a purely parallel approach minimizes latency at the cost of wasted compute on easy tasks, while a purely sequential design minimizes compute while creating high latency.
\emph{Iterative deepening} \citep{Korf1985IterativeDeepening} provides a mechanism to trade off latency and compute in a different way: assume parallel stages of exponentially growing sizes are conducted sequentially and the average pass rate is $\frac{1}{k}$. Then average latency is $\mathcal{O}(\log k)$ while compute is optimally $\mathcal{O}(k)$.

\begin{figure}[h]
    \centering
    \begin{subfigure}[b]{0.48\textwidth}
        \centering
        \includegraphics[width=\textwidth]{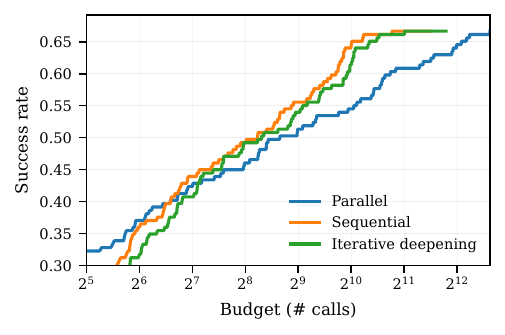}
        \caption{Compute}
        \label{fig:budget}
    \end{subfigure}
    \hfill
    \begin{subfigure}[b]{0.48\textwidth}
        \centering
        \includegraphics[width=\textwidth]{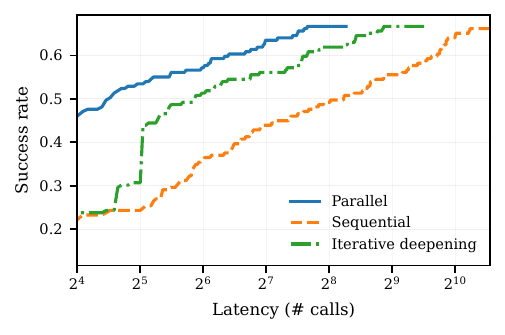}
        \caption{Latency}
        \label{fig:latency}
    \end{subfigure}
    \caption{\textbf{Performance by compute and latency.}}
    \label{fig:deepening}
\end{figure}

The subgoal decomposition system described in Section~\ref{sec:decomp} employs parallel provers, and we compare performance by latency and by compute for sequential, parallel and iterative deepening prover schedules. Independent goals are always conducted in parallel, while implementation subagents by design are always in sequence after a proving stage. Concretely, our experiment uses Claude 4.5 Opus on Verina with 8 provers per goal, and iterative deepening uses a $1 + 1 + 2 + 4$ schedule.  Figure~\ref{fig:deepening} shows how performance scales with compute and latency with the three sampling schemes. Because the first prover succeeds on many goals, iterative deepening approaches the performance of sequential scheduling in terms of total compute, while maintaining low latency that approaches it to the performance of parallel scheduling when comparing by latency. Note that this evaluation could be conducted post-hoc based on a single run with annotated prover agent launch order.

Based on these results, we decide to report multi-agent performance exclusively with iterative deepening.

\subsection{Model Comparison on Multi-Agent Inference Scaling}
\begin{figure}[h]
  \centering

  \begin{subfigure}{0.48\linewidth}
    \centering
    \includegraphics[width=\linewidth]{images/verina_decomp_models.pdf}
    \caption{Verina by compute}
    \label{fig:verina-decomp-budget}
  \end{subfigure}\hfill
  \begin{subfigure}{0.48\linewidth}
    \centering
    \includegraphics[width=\linewidth]{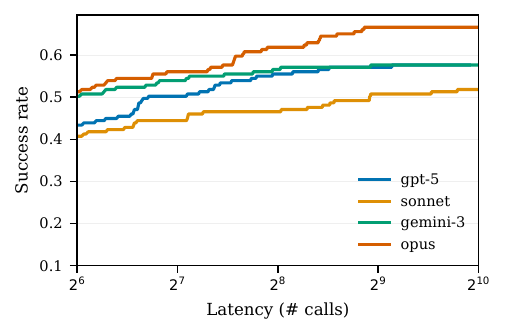}
    \caption{Verina by latency}
    \label{fig:verina-decomp-latency}
  \end{subfigure}

  \vspace{0.6em}

  \begin{subfigure}{0.48\linewidth}
    \centering
    \includegraphics[width=\linewidth]{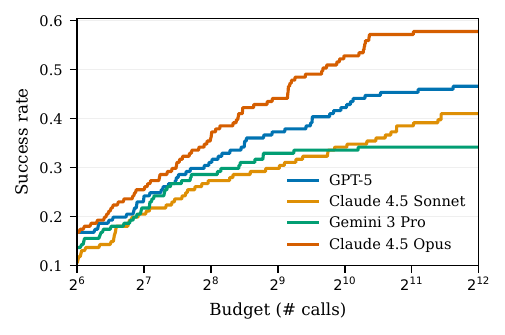}
    \caption{Clever by compute}
    \label{fig:clever-decomp-budget}
  \end{subfigure}\hfill
  \begin{subfigure}{0.48\linewidth}
    \centering
    \includegraphics[width=\linewidth]{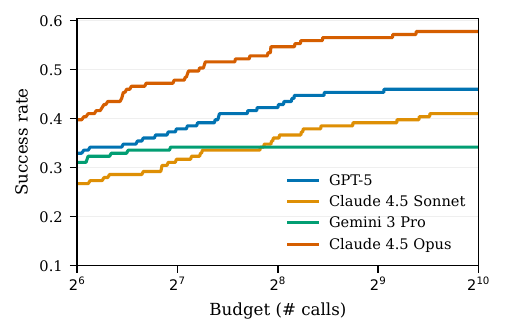}
    \caption{Clever by latency}
    \label{fig:clever-decomp-latency}
  \end{subfigure}

    \caption{\textbf{Inference scaling comparison for multi-agent system with different models.}
    We evaluate using up to 128 subagents on Verina and Clever and plot by the total budget of LLM calls spent and LLM call latency according to iterative deepening (Section~\ref{app:tradeoff}).
    }
  \label{fig:decomp-scale}
\end{figure}

Figure~\ref{fig:decomp-scale} shows how performance scales with respect to total call budget and latency, respectively, for each model.

\subsection{Scalability of Multi-Agent Systems}
\label{app:multi-atk}
One reason for scaling inference with multi-agent systems as opposed to pure independent pass@k scaling of single-agent systems is that by a carefully designed way of sharing state, individual subsystems can mutually inform and benefit each other. For an ideal multi-agent system, additional compute would always optimally be used to extend the resources of a single copy of the system, not to spawn multiple independent copies of it in pass@k fashion.

To evaluate the extent to which this applies to the subgoal decomposition multi-agent system from Section~\ref{sec:decomp}, we evaluate it using GPT-5 with $k=2$ copies on Clever. As can be seen in Figure~\ref{fig:decomp_at2}, increasing the compute allocated to a single copy of the system remains advantageous until a large budget of around 1200 calls is reached. The corresponding crossover point for sequential agents is at around 22.

\subsection{Sequential vs Parallel Compute}
\begin{figure}[h]
  \centering

  \begin{subfigure}{0.4\linewidth}
    \centering
    \includegraphics[width=\linewidth]{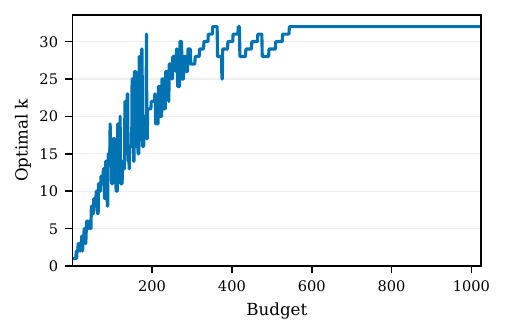}
    \caption{Optimal $k$ without tools}
    \label{fig:optimal-k}
  \end{subfigure}\hfill
  \begin{subfigure}{0.4\linewidth}
    \centering
    \includegraphics[width=\linewidth]{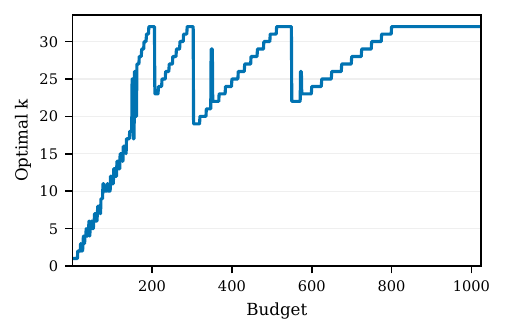}
    \caption{Optimal $k$ with tools}
    \label{fig:optimal-k-mcp}
  \end{subfigure}

  \vspace{0.6em}

  \begin{subfigure}{0.4\linewidth}
    \centering
    \includegraphics[width=\linewidth]{images/optimal_turns.pdf}
    \caption{Optimal $T$ without tools}
    \label{fig:optimal-t}
  \end{subfigure}\hfill
  \begin{subfigure}{0.4\linewidth}
    \centering
    \includegraphics[width=\linewidth]{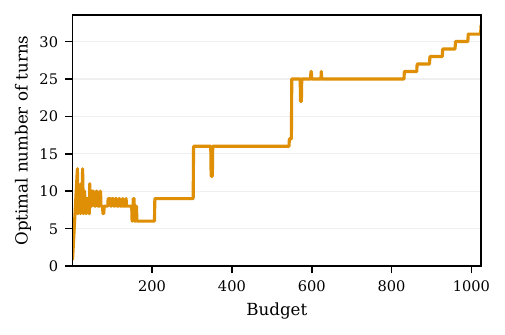}
    \caption{Optimal $T$ with tools}
    \label{fig:optimal-t-mcp}
  \end{subfigure}

  \caption{\textbf{Optimal inference budget allocation for GPT-5.} Given a maximum inference budget of up to $C$ language model calls, we compute the optimal breakdown into $kT \leq C$ with $k \leq 32$ the number of independent attempts and $T \leq 32$ the maximum number of turns per attempt.
  Optimal $T$ is larger than 1, meaning that multi-turn interactions are useful. However, $T$ saturates more quickly, suggesting that our bound $k \leq 32$ was suboptimal for a maximal budget of $1024 = 32^2$.
  For tool calling, the optimal $T$ has plateaus at discrete levels of approximately 8, 16 and 24 turns, respectively. This could be an artifact of discrete post-training recipes, with models optimizing their exploration/exploitation tradeoff based on an inferred maximum number of turns.}
  \label{fig:optimal}
\end{figure}

For sequential agents, there are two ways of increasing the inference compute budget: by increasing the maximum number of turns $T$ or the number of independent attempts $k$. In Figure~\ref{fig:optimal}, we evaluate with $k_\mathrm{max}=32, T_\mathrm{max}=32$ and compute from this data the optimal $k$ and $T$ for varying compute budgets $C \geq kT$. The results indicate that for GPT-5 on Clever, multi-turn iteration is useful at moderate numbers of iterations, and that boundary effects appear for the tool use agent where iteration lengths of 8, 16 and 24 have particular importance.

\section{Extended Qualitative Evaluation}
\label{app:quali}

To better understand the current limitations of WybeCoder, we considered a number of classic sorting algorithms; see Table~\ref{tab:sorts}.
We prompted asking for well known algorithms and for proofs of them. The first four passed without difficulty. The next three sorts are more advanced and are often presented recursively. Loom currently has limited support for recursion, so we asked for iterative variants. Mergesort is bottom up with $\mathcal{O}(n)$ extra memory, while Heapsort and Quicksort are standard in-place iterative versions. To ease the verification effort, we manually provided specifications of several sub-routines and not only of the top-level main functions. 

The verifications of the main functions themselves all went through, based on these specs. In Quicksort and Mergesort, the proofs of the most challenging sub-procedures \texttt{mergepass} and \texttt{quickstep} failed. 

We describe our experience with Heapsort because of lessons we learned from it. We began by manually providing specification for the three sub-routines \texttt{heapify}, \texttt{maxheap} and \texttt{heapsort\_main} with LLM assistance. The three procedures were then generated and proven. Although the overall specification of sorting was satisfied, human examination of the proof script revealed a curious case. In a comment in \texttt{heapify}, the model wrote:
\begin{lstlisting}[language=lean, style=lean]
-- Since Heapify may permute elements outside the heap, we cannot guarantee
-- sortedness from the heapsort phase. Apply selection sort to ensure sortedness.
\end{lstlisting}

The \texttt{heapsort\_main} routine included a selection sort within it! The selection sort code was in fact unnecessary, because \texttt{heapify} does not permute outside the heap. The technical difficulty was that the manually provided specification for \texttt{heapify} left out a \emph{frame axiom}, saying what does not change \cite{mccarthy1969philosophical}. Once we corrected the spec to include the missing frame axiom, a correct version of Heapsort was generated, and the verification went through.

This experience illustrates several points on the gap from current capabilities to a verified form of vibe coding which is effective at scale: (i) humans (with LLM help) wrote the specs for the sub-procedures, where ideally the AI would do so; (ii) neither humans nor LLM spotted the missing frame axiom, or the fact that it could be consistently added to the spec based on the code; and (iii) the AI valued task completion higher than abiding by implictit intention and did not object to the version with embedded selection sort, a human was needed.

All told, these points indicate how proficiency in doing proofs of individual procedures shifts the bottleneck towards \emph{specification engineering and review}. This mirrors a development in machine learning for formal mathematics where increased proving capabilities move the bottleneck toward \emph{formalizing definitions} correctly and in line with the practical requirements of a given interactive theorem prover.

\section{Examples}
\label{app:examples}

\subsection{Heapsort}
\label{appendix:heapsort}

To evaluate the limits of the WybeCoder framework's ability to synthesize more complex algorithms, we specified heapsort by decomposing it into three methods with formal pre- and post-conditions:

\begin{enumerate}
    \item \texttt{Heapify\_ArrayPerm}: The sift-down operation that restores the max-heap property at a given index, assuming the subtrees rooted at its children already satisfy the heap property.
    
    \item \texttt{BuildMaxHeap\_ArrayPerm}: Constructs a max-heap from an arbitrary array by iteratively calling \texttt{Heapify\_ArrayPerm} from the last non-leaf node down to the root.
    
    \item \texttt{HeapSort\_ArrayPerm}: The main sorting algorithm that first builds a max-heap, then repeatedly extracts the maximum element by swapping it to the end and restoring the heap property on the reduced prefix.
\end{enumerate}

Each method was specified with formal contracts ensuring:
\begin{itemize}
    \item \textbf{Size preservation}: The output array has the same size as the input.
    \item \textbf{Permutation invariance}: The output is a permutation of the input (\texttt{List.Perm arr.toList result.toList}).
    \item \textbf{Correctness properties}: \texttt{Heapify\_ArrayPerm} establishes the heap property at the target index; \texttt{BuildMaxHeap\_ArrayPerm} produces a valid max-heap; \texttt{HeapSort\_ArrayPerm} produces a sorted array (\texttt{List.Sorted ($\cdot \leq \cdot$) result.toList}).
    \item \textbf{Frame conditions}: Elements outside the active heap region remain unchanged.
\end{itemize}

Using Claude Opus 4.5, the framework successfully generated both the executable Lean implementations and complete correctness proofs for all three methods. The resulting verified implementation spans approximately 2,000 lines of Lean~4 code, including loop invariants, auxiliary lemmas about heap index arithmetic (e.g., \texttt{parent}, \texttt{left}, \texttt{right} relationships), and the main correctness theorems. The proof obligations---particularly the loop invariant preservation for \texttt{Heapify\_ArrayPerm}'s sift-down loop and the partition/sortedness invariants for \texttt{HeapSort\_ArrayPerm}---required substantial automated reasoning about array permutations, integer arithmetic, and the recursive structure of the heap property.

This case study demonstrates that WybeCoder can automate verification work that would otherwise require multiple days of effort from a team of expert formal methods practitioners, while producing machine-checked proofs that provide strong correctness guarantees. We note that the proving pipeline does not optimize proof length, maintainability or elegance, and that a minimal proof might be significantly shorter.

\lstinputlisting[caption={Implementation and correctness proof of Heapsort for array inputs},label={lst:heapsort}, captionpos=t, language=lean, breaklines=true, style=lean]
{listings/heapsort.lean}

\subsection{Prime Factorization}
\label{appendix:factorize}

WybeCoder using Clause Opus 4.5 successfully verified a prime factorization algorithm that decomposes a natural number $n \geq 2$ into its prime factors with multiplicity. This is problem \texttt{clever\_id\_24} from the Clever dataset. The algorithm maintains a trial divisor \texttt{factor} (starting at 2) and repeatedly divides out factors from \texttt{remaining} until the quotient becomes 1. The termination proof of the \texttt{while} loop uses the non-standard decreasing measure $2 \cdot \texttt{remaining} - \texttt{factor}$. In the division branch, \texttt{remaining} strictly decreases; in the increment branch, \texttt{factor} increases while \texttt{remaining} stays fixed. Proving the measure decreases in the latter case requires establishing that \texttt{factor} $\leq$ \texttt{remaining}, a non-trivial fact following from the invariant that some prime must divide the remaining quotient.
The complete verified implementation spans approximately 200 lines of Lean~4 code, including five helper lemmas for sorted list extension, product preservation under division, and primality/divisibility propagation. This case study demonstrates WybeCoder's ability to handle algorithms requiring non-trivial termination arguments and interdependent loop invariants.

\begin{lstlisting}[caption={Implementation and correctness proof of a Prime Factorization (\texttt{clever\_id\_24})},label={lst:factorize}, captionpos=t, language=lean, breaklines=true, style=lean]
import Auto
import Aesop
import Lean
import Mathlib

import CaseStudies.Velvet.Std
import CaseStudies.TestingUtil

set_option loom.semantics.termination "total"
set_option loom.semantics.choice "demonic"
set_option loom.solver "cvc5"
set_option auto.smt.timeout 3
set_option maxHeartbeats 100000
set_option auto.smt.trust true


method factorize (n: ℕ) return (res: List ℕ)
  require 2 ≤ n
  ensures res.prod = n
  ensures List.Sorted (· ≤ ·) res
  ensures ∀ i ∈ res, n % i = 0 ∧ Nat.Prime i
  do
    let n₀ := n
    let mut remaining := n
    let mut factor := 2
    let mut factors : List ℕ := []

    while remaining > 1
      invariant remaining ≥ 1
      invariant factor ≥ 2
      invariant factors.prod * remaining = n₀
      invariant ∀ p ∈ factors, Nat.Prime p ∧ n₀ % p = 0
      invariant List.Sorted (· ≤ ·) factors
      invariant ∀ p ∈ factors, p ≤ factor
      invariant ∀ p, Nat.Prime p → p < factor → remaining % p ≠ 0
      decreasing remaining + remaining - factor
    do
      if remaining % factor = 0 then
        factors := factors ++ [factor]
        remaining := remaining / factor
      else
        factor := factor + 1

    return factors

-- Helper lemma for sorted append
theorem sorted_append_singleton {l : List ℕ} {x : ℕ} 
    (hs : List.Sorted (· ≤ ·) l) (hx : ∀ y ∈ l, y ≤ x) : 
    List.Sorted (· ≤ ·) (l ++ [x]) := by
  induction l with
  | nil => simp [List.Sorted]
  | cons head tail ih =>
    simp only [List.cons_append]
    rw [List.sorted_cons] at hs ⊢
    constructor
    · intro b hb
      simp only [List.mem_append, List.mem_singleton] at hb
      cases hb with
      | inl ht => exact hs.1 b ht
      | inr hx' => 
        rw [hx']
        have hmem : head ∈ head :: tail := by simp
        exact hx head hmem
    · apply ih hs.2
      intro y hy
      have hmem : y ∈ head :: tail := List.mem_cons_of_mem head hy
      exact hx y hmem

-- Helper: factor is prime when it divides remaining and no smaller prime does
theorem factor_is_prime (remaining factor : ℕ) 
    (hrem : remaining > 1) (hfac : factor ≥ 2) 
    (hdiv : factor ∣ remaining)
    (hno_smaller : ∀ p, Nat.Prime p → p < factor → ¬(p ∣ remaining)) :
    Nat.Prime factor := by
  have hne1 : remaining ≠ 1 := by omega
  have hmin_le : remaining.minFac ≤ factor := Nat.minFac_le_of_dvd hfac hdiv
  have hmin_prime := Nat.minFac_prime hne1
  have hmin_dvd := Nat.minFac_dvd remaining
  by_cases heq : remaining.minFac = factor
  · rw [← heq]; exact hmin_prime
  · have hlt : remaining.minFac < factor := Nat.lt_of_le_of_ne hmin_le heq
    exfalso
    exact hno_smaller remaining.minFac hmin_prime hlt hmin_dvd

-- Helper for product preservation
theorem prod_append_div (factors : List ℕ) (remaining factor n : ℕ) 
    (hdiv : factor ∣ remaining) (hprod : factors.prod * remaining = n) :
    (factors ++ [factor]).prod * (remaining / factor) = n := by
  simp only [List.prod_append, List.prod_singleton]
  have heq : remaining / factor * factor = remaining := Nat.div_mul_cancel hdiv
  calc factors.prod * factor * (remaining / factor) 
      = factors.prod * (remaining / factor * factor) := by ring
    _ = factors.prod * remaining := by rw [heq]
    _ = n := hprod

-- Helper for primality in append
theorem prime_of_mem_append_factor {p factor : ℕ} {factors : List ℕ} {remaining n : ℕ}
    (h : p ∈ factors ++ [factor])
    (hfac_ge : factor ≥ 2) (hrem_gt : remaining > 1)
    (hmod : remaining % factor = 0)
    (hinv : ∀ q ∈ factors, Nat.Prime q ∧ n % q = 0)
    (hno_smaller : ∀ q, Nat.Prime q → q < factor → remaining % q ≠ 0) :
    Nat.Prime p := by
  simp only [List.mem_append, List.mem_singleton] at h
  rcases h with hfactors | heq
  · exact (hinv p hfactors).1
  · rw [heq]
    have hdiv : factor ∣ remaining := Nat.dvd_of_mod_eq_zero hmod
    have hno_smaller' : ∀ q, Nat.Prime q → q < factor → ¬(q ∣ remaining) := by
      intro q hq hlt hdvd
      have := hno_smaller q hq hlt
      have : remaining % q = 0 := Nat.mod_eq_zero_of_dvd hdvd
      contradiction
    exact factor_is_prime remaining factor hrem_gt hfac_ge hdiv hno_smaller'

-- Helper for divisibility in append
theorem div_of_mem_append_factor {p factor : ℕ} {factors : List ℕ} {remaining n : ℕ}
    (h : p ∈ factors ++ [factor])
    (hmod : remaining % factor = 0)
    (hprod : factors.prod * remaining = n)
    (hinv : ∀ q ∈ factors, Nat.Prime q ∧ n % q = 0) :
    n % p = 0 := by
  simp only [List.mem_append, List.mem_singleton] at h
  rcases h with hfactors | heq
  · exact (hinv p hfactors).2
  · rw [heq]
    have hdiv : factor ∣ remaining := Nat.dvd_of_mod_eq_zero hmod
    have hrem_dvd_n : remaining ∣ n := by use factors.prod; linarith
    have hdiv_n : factor ∣ n := Nat.dvd_trans hdiv hrem_dvd_n
    exact Nat.mod_eq_zero_of_dvd hdiv_n

-- Solver hints
attribute [local solverHint] Nat.div_lt_self
attribute [local solverHint] Nat.Prime.two_le
attribute [local solverHint] List.prod_append
attribute [local solverHint] List.prod_singleton

set_option maxHeartbeats 4000000 in
set_option auto.smt.timeout 10 in
prove_correct factorize by
  loom_solve <;> try grind
  all_goals try simp only [List.prod_nil, List.Sorted, List.Pairwise.nil, one_mul]
  -- Handle product preservation
  all_goals try (
    have hdiv : factor ∣ remaining := Nat.dvd_of_mod_eq_zero ‹remaining % factor = 0›
    have hprod : factors.prod * remaining = n := ‹factors.prod * remaining = n›
    exact prod_append_div factors remaining factor n hdiv hprod
  )
  -- Handle remaining / factor ≥ 1
  all_goals try (
    have hdiv : factor ∣ remaining := Nat.dvd_of_mod_eq_zero ‹remaining % factor = 0›
    have hfac_pos : factor > 0 := by omega
    have hrem_ge : remaining ≥ factor := Nat.le_of_dvd (by omega) hdiv
    have := Nat.div_pos hrem_ge hfac_pos
    omega
  )
  -- Handle sorted append
  all_goals try (
    apply sorted_append_singleton ‹List.Sorted (· ≤ ·) factors›
    exact ‹∀ p ∈ factors, p ≤ factor›
  )
  -- Handle decreasing measure for if_pos branch  
  all_goals try (
    have hdiv : factor ∣ remaining := Nat.dvd_of_mod_eq_zero ‹remaining % factor = 0›
    have hfac_ge : factor ≥ 2 := ‹factor ≥ 2›
    have hrem_gt : remaining > 1 := ‹remaining > 1›
    have hrem_ge : remaining ≥ factor := Nat.le_of_dvd (by omega) hdiv
    have hdiv_lt := Nat.div_lt_self (by omega : remaining > 0) (by omega : factor > 1)
    omega
  )
  -- Handle decreasing for else branch
  all_goals try (
    have hrem_gt : remaining > 1 := ‹remaining > 1›
    have hfac_ge : factor ≥ 2 := ‹factor ≥ 2›
    have hfac_le : factor ≤ remaining := by
      by_contra hgt
      push_neg at hgt
      have : remaining.minFac ≤ remaining := Nat.minFac_le (by omega)
      have hprime : Nat.Prime remaining.minFac := Nat.minFac_prime (by omega)
      have hdvd : remaining.minFac ∣ remaining := Nat.minFac_dvd remaining
      have hge2 : remaining.minFac ≥ 2 := Nat.Prime.two_le hprime
      have hlt : remaining.minFac < factor := by omega
      have := ‹∀ p, Nat.Prime p → p < factor → remaining % p ≠ 0› remaining.minFac hprime hlt
      have : remaining % remaining.minFac = 0 := Nat.mod_eq_zero_of_dvd hdvd
      contradiction
    omega
  )
  -- Handle Nat.Prime p for p ∈ factors ++ [factor]
  all_goals try (
    exact prime_of_mem_append_factor h ‹factor ≥ 2› ‹remaining > 1› ‹remaining % factor = 0› 
      ‹∀ p ∈ factors, Nat.Prime p ∧ n % p = 0› ‹∀ p, Nat.Prime p → p < factor → remaining % p ≠ 0›
  )
  -- Handle n % p = 0 for p ∈ factors ++ [factor]
  all_goals try (
    exact div_of_mem_append_factor h ‹remaining % factor = 0› ‹factors.prod * remaining = n› 
      ‹∀ p ∈ factors, Nat.Prime p ∧ n % p = 0›
  )
  -- Handle p ≤ factor for p ∈ factors ++ [factor]
  all_goals try (
    simp only [List.mem_append, List.mem_singleton] at h
    rcases h with hfactors | heq
    · exact ‹∀ p ∈ factors, p ≤ factor› p hfactors
    · rw [heq]
  )
  -- Handle remaining / factor % p ≠ 0 for primes p < factor
  all_goals try (
    have hdiv : factor ∣ remaining := Nat.dvd_of_mod_eq_zero ‹remaining % factor = 0›
    have hno := ‹∀ q, Nat.Prime q → q < factor → remaining % q ≠ 0› p ‹Nat.Prime p› ‹p < factor›
    intro hdivp
    have hdvd : p ∣ remaining / factor := Nat.dvd_of_mod_eq_zero hdivp
    have hdvd2 : p ∣ remaining := Nat.dvd_trans hdvd (Nat.div_dvd_of_dvd hdiv)
    have hmod0 : remaining % p = 0 := Nat.mod_eq_zero_of_dvd hdvd2
    contradiction
  )

#print axioms factorize_correct
\end{lstlisting}

\subsection{Example: Kadane's algorithm}
To give a full example of a standard algorithm that was verified by WybeCoder using GPT-5 using the subgoal decomposition pipeline described in Section~\ref{sec:linear}, consider the maximum subarray problem: Given a list of numbers, the goal is to find the maximal sum of a contiguous sublist. Kadane's algorithm solves this task in $\mathcal{O}(n)$ time complexity. It is defined as follows:

\begin{lstlisting}[caption={Kadane's algorithm}, style=python]
def max_subarray(numbers):
    best = 0
    current = 0
    for x in numbers:
        current = max(0, current + x)
        best = max(best, current)
    return best
\end{lstlisting}
For the correctness proof, we argue as follows. The variable \texttt{current} holds the maximal subarray sum for subarrays that end at the index before \texttt{x}. The variable \texttt{best} holds the maximal subarray sum for the subarray that ends before \texttt{x}, i.e. the running maximum over the values of \texttt{current} with the straightforward update rule. The update rule for \texttt{current} considers two cases: If the maximal subarray contains \texttt{x}, it can be extended to the left using the maximal subarray that ends at the position before \texttt{x}. Otherwise, it is the empty subarray and \texttt{current} must be set to zero. Which of the two happens to be the case depends on whether \texttt{current + x} is positive or not. We leave the details to the reader to appreciate the complexity of writing a complete formal argument.

Listing~\ref{lst:kadane} shows the full formal proof obtained with our pipeline. The method implementation uses \emph{ghost variables} such as \texttt{curStart} which are only required to formulate the invariants as hinted at by the above informal argument. The actual algorithm matches the pseudo-code above exactly, while there is significant ``formalization overhead'' from invariant annotations and ghost variables updates. The proof consists of several lemmas generated by the prover agents (one of which for dealing with Loom/Velvet internals), followed by the main correctness theorem with the \texttt{loom\_solve} tactic for generating verification conditions and discharging them to CVC5, and the reassembled subgoal proofs obtained from the subgoal proof agents. We note that the proving pipeline does not optimize proof length, maintainability or elegance, and that a minimal proof might be significantly shorter.


\begin{lstlisting}[caption={Implementation and correctness proof for Kadane's algorithm (\texttt{verina\_advanced\_46})},label={lst:kadane}, captionpos=t, language=lean, breaklines=true, style=lean]
import Auto
import Lean
import Mathlib

import CaseStudies.Velvet.Std
import CaseStudies.TestingUtil

set_option loom.semantics.termination "total"
set_option loom.semantics.choice "demonic"
set_option loom.solver "cvc5"
set_option auto.smt.timeout 3
set_option maxHeartbeats 100000
set_option auto.smt.trust true



method maxSubarraySum (numbers : List Int) return (result : Int)
  ensures let subArraySums :=
      List.range (numbers.length + 1) |>.flatMap (fun start =>
        List.range (numbers.length - start + 1) |>.map (fun len =>
          numbers.drop start |>.take len |>.sum))
    subArraySums.contains result ∧ subArraySums.all (· ≤ result)
  do
    let mut i : Nat := 0
    let mut rest : List Int := numbers
    let mut cur : Int := 0
    let mut max : Int := 0
    let mut curStart : Nat := 0
    let mut bestStart : Nat := 0
    let mut bestLen : Nat := 0
    while rest ≠ []
      invariant hi_len : i + rest.length = numbers.length
      invariant hrest_eq : rest = List.drop i numbers
      invariant hi_bounds : 0 ≤ i ∧ i ≤ numbers.length
      invariant hcurStart_le : curStart ≤ i
      invariant hcur_nonneg : 0 ≤ cur
      invariant hcur_sum : cur = (List.take (i - curStart) (List.drop curStart numbers)).sum
      invariant hcur_suffix_max :
        forall (start : Nat),
          start ≤ i →
          (List.take (i - start) (List.drop start numbers)).sum ≤ cur
      invariant hmax_nonneg : 0 ≤ max
      invariant hbest_sum : (List.take bestLen (List.drop bestStart numbers)).sum = max
      invariant hbest_start_in_prefix : bestStart ≤ i
      invariant hbest_end_in_prefix : bestStart + bestLen ≤ i
      invariant hprefix_max :
        forall (start : Nat) (len : Nat),
          start ≤ i ∧ start + len ≤ i →
          (List.take len (List.drop start numbers)).sum ≤ max
      done_with hdone : rest = []
      decreasing d : rest.length
    do
      let x := rest.head!
      let tail := rest.tail
      if cur + x ≤ 0 then
        cur := 0
        curStart := i + 1
      else
        cur := cur + x
      if max < cur then
        max := cur
        bestStart := curStart
        bestLen := (i + 1) - curStart
      i := i + 1
      rest := tail
    return max


@[simp] lemma WithName_erase_length {α} (l : List α) (n : Lean.Name) :
  (WithName.mk' l n).erase.length = l.length := by
  rfl

lemma take_succ_eq_append_head! {α} [Inhabited α]
    (m : List α) (k : Nat) (h : List.drop k m ≠ []) :
  List.take (k+1) m = List.take k m ++ [ (List.drop k m).head! ] := by
  revert m
  induction k with
  | zero =>
    intro m h
    cases m with
    | nil =>
      simp at h
    | cons x xs =>
      simp [List.drop, List.take, List.head!]
  | succ k ih =>
    intro m h
    cases m with
    | nil =>
      simp at h
    | cons x xs =>
      have hx : List.drop k xs ≠ [] := by
        simpa [List.drop] using h
      have ih' := ih xs hx
      simpa [List.drop, List.take, ih', List.cons_append]

lemma sum_take_succ_eq_sum_take_add_head! {α} [Inhabited α] [AddCommMonoid α]
  (l : List α) (n : Nat) (h : l.drop n ≠ []) :
  (l.take (n + 1)).sum = (l.take n).sum + (l.drop n).head! := by
  have hrepr := take_succ_eq_append_head! l n h
  calc
    (List.take (n+1) l).sum
        = (List.take n l ++ [ (List.drop n l).head! ]).sum := by simpa [hrepr]
    _ = (List.take n l).sum + [ (List.drop n l).head! ].sum := by
          simpa [List.sum_append]
    _ = (List.take n l).sum + (List.drop n l).head! := by simp

lemma sum_take_one_head! (l : List ℤ) (h : l ≠ []) : (List.take 1 l).sum = l.head! := by
  cases l with
  | nil => cases h rfl
  | cons a t => simp [List.take]

lemma sum_take_succ_of_drop_ne_nil (l : List ℤ) (n : ℕ) (hne : List.drop n l ≠ []) :
  (List.take (Nat.succ n) l).sum = (List.take n l).sum + (List.drop n l).head! := by
  revert l
  induction n with
  | zero =>
    intro l hne
    cases l with
    | nil =>
      cases hne rfl
    | cons x xs =>
      simp [List.take, List.drop, List.sum_cons, List.head!]
  | succ n ih =>
    intro l hne
    cases l with
    | nil =>
      simp at hne
    | cons x xs =>
      have hne' : List.drop n xs ≠ [] := by
        simpa [List.drop] using hne
      have ih' := ih xs hne'
      simp [List.take, List.drop, List.sum_cons, ih', add_comm, add_left_comm, add_assoc]

set_option maxHeartbeats 1000000 in
prove_correct maxSubarraySum by
  loom_solve
  case hi_len.loop_1 => (
  rcases List.exists_cons_of_ne_nil (l := rest) if_pos with ⟨x, xs, hrest⟩
  have hLHS :
      i + 1 + (WithName.mk' rest.tail (Lean.Name.anonymous.mkStr "rest")).erase.length
        = i + 1 + xs.length := by
    simp [WithName_erase_length, hrest]
  have hlen :
      i + xs.length + 1 = numbers.length := by
    simpa [hrest, Nat.add_comm, Nat.add_left_comm, Nat.add_assoc] using hi_len
  have hreassoc : i + 1 + xs.length = i + xs.length + 1 := by
    simpa [Nat.add_comm, Nat.add_left_comm, Nat.add_assoc]
  exact by
    rw [hLHS, hreassoc, hlen]
  )
  case hrest_eq.loop_1 => (
  try simp_all ; try grind
  )
  case hi_bounds.loop_3 => (
  cases rest with
  | nil =>
      cases if_pos rfl
  | cons x tail =>
      have hlen_pos : 0 < (x :: tail).length := by simp
      have hstep : i + 1 ≤ i + (x :: tail).length :=
        Nat.add_le_add_left (Nat.succ_le_of_lt hlen_pos) i
      simpa using (hi_len ▸ hstep)
  )
  case hcur_sum.loop_1 => (
  try simp_all ; try grind
  )
  case hcur_suffix_max.loop_1 => (
  classical
  by_cases hstart : start = i + 1
  · subst hstart
    simp
  ·
    have hlt : start < i + 1 := lt_of_le_of_ne a_2 hstart
    have hle_start_i : start ≤ i := Nat.le_of_lt_succ hlt
    have eqi : start + (i - start) = i := by
      simpa [Nat.add_comm] using (Nat.sub_add_cancel hle_start_i)
    have hlen_eq : i + 1 - start = (i - start) + 1 := by
      calc
        i + 1 - start
            = (start + (i - start) + 1) - start := by
                simpa [eqi, Nat.add_comm, Nat.add_left_comm, Nat.add_assoc]
        _ = ((i - start) + 1) := by
                simpa [Nat.add_comm, Nat.add_left_comm, Nat.add_assoc]
                  using (Nat.add_sub_cancel start ((i - start) + 1))
    have hdecomp :
        List.take (i + 1 - start) (List.drop start numbers)
          = List.take (i - start) (List.drop start numbers)
            ++ List.take 1 (List.drop i numbers) := by
      have := (List.take_add (l := List.drop start numbers) (i := i - start) (j := 1))
      simpa [hlen_eq, List.drop_drop, eqi, Nat.add_comm, Nat.add_left_comm, Nat.add_assoc] using this
    have hsum_eq :
        (List.take (i + 1 - start) (List.drop start numbers)).sum
          = (List.take (i - start) (List.drop start numbers)).sum
            + (List.take 1 (List.drop i numbers)).sum := by
      calc
        _ = ((List.take (i - start) (List.drop start numbers)) ++ List.take 1 (List.drop i numbers)).sum := by
              simp [hdecomp]
        _ = (List.take (i - start) (List.drop start numbers)).sum
              + (List.take 1 (List.drop i numbers)).sum := by
              simpa using (List.sum_append
                (List.take (i - start) (List.drop start numbers))
                (List.take 1 (List.drop i numbers)))
    have hsum_take1 :
        (List.take 1 (List.drop i numbers)).sum = (List.drop i numbers).head! := by
      have : (List.take 1 rest).sum = rest.head! := by
        cases rest <;> simp
      simpa [hrest_eq] using this
    have hprev_le_cur :
        (List.take (i - start) (List.drop start numbers)).sum ≤ cur :=
      hcur_suffix_max start hle_start_i
    have hcur_head_le0 :
        cur + (List.drop i numbers).head! ≤ 0 := by
      simpa [hrest_eq] using if_pos_1
    calc
      (List.take (i + 1 - start) (List.drop start numbers)).sum
          = (List.take (i - start) (List.drop start numbers)).sum
              + (List.take 1 (List.drop i numbers)).sum := hsum_eq
      _ = (List.take (i - start) (List.drop start numbers)).sum
              + (List.drop i numbers).head! := by simpa [hsum_take1]
      _ ≤ cur + (List.drop i numbers).head! := by
            exact add_le_add_right hprev_le_cur _
      _ ≤ 0 := hcur_head_le0
  )
  case hprefix_max.loop_1 => (
  classical
  cases Nat.lt_or_eq_of_le a_3 with
  | inl hlt =>
    have hle_i : start + len ≤ i := Nat.lt_succ_iff.mp hlt
    have hstart_le_i : start ≤ i := Nat.le_trans (Nat.le_add_right _ _) hle_i
    exact hprefix_max start len hstart_le_i hle_i
  | inr heq =>
    cases Nat.eq_or_lt_of_le a_2 with
    | inl hstart_eq =>
      have : (i + 1) + len = i + 1 := by simpa [hstart_eq] using heq
      have hlen0 : len = 0 := Nat.add_left_cancel this
      simp [hlen0, hmax_nonneg]
    | inr hstart_lt =>
      have hstart_le_i : start ≤ i := Nat.lt_succ_iff.mp hstart_lt
      have hlen_sub : (i + 1) - start = len := by
        have := congrArg (fun t => t - start) heq.symm
        simpa [Nat.add_sub_cancel] using this
      have hlen_succ : (i + 1) - start = (i - start) + 1 := by
        have hi_eq : start + (i - start) = i := Nat.add_sub_of_le hstart_le_i
        calc
          (i + 1) - start
              = ((start + (i - start)) + 1) - start := by simpa [hi_eq]
          _ = (start + ((i - start) + 1)) - start := by
                simpa [Nat.add_assoc]
          _ = (i - start) + 1 := by
                simpa using (Nat.add_sub_cancel start ((i - start) + 1))
      have hsum_decomp :
          (List.take len (List.drop start numbers)).sum
            =
          (List.take (i - start) (List.drop start numbers)).sum
            +
          (List.take 1 (List.drop i numbers)).sum := by
        calc
          (List.take len (List.drop start numbers)).sum
              = (List.take ((i + 1) - start) (List.drop start numbers)).sum := by
                  simpa [hlen_sub]
          _ = (List.take ((i - start) + 1) (List.drop start numbers)).sum := by
                  simpa [hlen_succ]
          _ = (List.take (i - start) (List.drop start numbers)).sum
                + (List.take 1 (List.drop (i - start) (List.drop start numbers))).sum := by
                  simp [List.take_add, List.sum_append]
          _ = (List.take (i - start) (List.drop start numbers)).sum
                + (List.take 1 (List.drop i numbers)).sum := by
                  have hi_eq : start + (i - start) = i := Nat.add_sub_of_le hstart_le_i
                  have h1 := List.drop_drop (l := numbers) (i := i - start) (j := start)
                  have hdrop_eq :
                      List.drop (i - start) (List.drop start numbers) = List.drop i numbers := by
                    simpa [Nat.add_comm, hi_eq] using h1
                  simpa [hdrop_eq]
      have hdrop_ne_nil : List.drop i numbers ≠ [] := by simpa [hrest_eq] using if_pos
      have htake1_eq_head : (List.take 1 (List.drop i numbers)).sum = (List.drop i numbers).head! :=
        sum_take_one_head! _ hdrop_ne_nil
      have hold_le_cur :
          (List.take (i - start) (List.drop start numbers)).sum ≤ cur :=
        hcur_suffix_max start hstart_le_i
      have hsum_le_zero :
          (List.take (i - start) (List.drop start numbers)).sum
            + (List.drop i numbers).head!
            ≤ 0 := by
        have h1 :
            (List.take (i - start) (List.drop start numbers)).sum
              + (List.drop i numbers).head!
              ≤
            cur + (List.drop i numbers).head! :=
          add_le_add_right hold_le_cur _
        have h2 : cur + (List.drop i numbers).head! ≤ 0 := by
          simpa [hrest_eq, htake1_eq_head] using if_pos_1
        exact le_trans h1 h2
      have hnonneg_max : 0 ≤ max := hmax_nonneg
      have hsum_le_max :
          (List.take (i - start) (List.drop start numbers)).sum
            + (List.drop i numbers).head!
            ≤ max :=
        le_trans hsum_le_zero hnonneg_max
      simpa [hsum_decomp, htake1_eq_head]
        using hsum_le_max
  )
  case decreasing.loop_1 => (
  classical
  cases rest with
  | nil =>
      cases if_pos rfl
  | cons x xs =>
      change ((WithName.mk' xs (Lean.Name.anonymous.mkStr "rest")).erase.length) < Nat.succ xs.length
      have h : (WithName.mk' xs (Lean.Name.anonymous.mkStr "rest")).erase = xs := rfl
      simpa [h] using Nat.lt_succ_self xs.length
  )
  case hi_len.loop_2 => (
  cases rest with
  | nil =>
    cases if_pos rfl
  | cons x xs =>
    have h_erase :
        (WithName.mk' ((x :: xs).tail) (Lean.Name.anonymous.mkStr "rest")).erase
        = (x :: xs).tail := rfl
    have hi_len' : i + (List.length (x :: xs)) = numbers.length := hi_len
    simpa [h_erase, List.tail, List.length, Nat.add_assoc, Nat.add_comm, Nat.add_left_comm] using hi_len'
  )
  case hrest_eq.loop_2 => (
  try simp_all ; try grind
  )
  case hi_bounds.loop_5 => (
  cases rest with
  | nil =>
    cases if_pos rfl
  | cons x xs =>
    have hi_len' : i + (xs.length + 1) = numbers.length := by
      simpa [List.length_cons, Nat.add_comm, Nat.add_assoc] using hi_len
    have h : i + 1 ≤ (i + 1) + xs.length := Nat.le_add_right (i + 1) xs.length
    have h' : i + 1 ≤ i + (xs.length + 1) := by
      simpa [Nat.add_assoc, Nat.add_comm] using h
    simpa [hi_len'] using h'
  )
  case hcur_sum.loop_2 => (
  classical
  have hrestne : List.drop i numbers ≠ [] := by simpa [hrest_eq] using if_pos
  set xs := List.drop curStart numbers
  set k := i - curStart
  have hdrop_eq :
      List.drop k xs = List.drop i numbers := by
    simpa [xs, k, Nat.add_sub_of_le hcurStart_le] using
      (List.drop_drop (l := numbers) (n := curStart) (m := i - curStart))
  have hdrop_ne : List.drop k xs ≠ [] := by
    intro h
    exact hrestne (by simpa [hdrop_eq] using h)
  cases h : List.drop k xs with
  | nil =>
      cases hdrop_ne (by simpa [h])
  | cons y ys =>
      have hcons : List.drop k xs = y :: ys := h
      have hkadd : i + 1 - curStart = k + 1 := by
        have h' : curStart + (i - curStart) = i := Nat.add_sub_of_le hcurStart_le
        have step1 :
          i + 1 - curStart = (curStart + (i - curStart) + 1) - curStart := by
          simpa [Nat.add_assoc] using (congrArg (fun t => t + 1 - curStart) h').symm
        have step2 :
          (curStart + (i - curStart) + 1) - curStart
            = (curStart + ((i - curStart) + 1)) - curStart := by
          simpa [Nat.add_assoc]
        have step3 :
          (curStart + ((i - curStart) + 1)) - curStart
            = (i - curStart) + 1 := by
          simpa using (Nat.add_sub_cancel curStart ((i - curStart) + 1))
        exact by
          simpa [k] using (step1.trans (step2.trans step3))
      have hhead' :
        (List.drop k xs).head! = y := by
        simp [hcons]
      have hhead :
        (List.drop i numbers).head! = y := by
        simpa [hdrop_eq] using hhead'
      have h1 :
        ((List.take k xs).sum) + (List.drop i numbers).head!
          = ((List.take k xs).sum) + y := by
        simpa [hhead]
      have h2 :
        ((List.take k xs).sum) + y
          = ((List.take k xs) ++ [y]).sum := by
        simpa using (List.sum_append (l₁ := List.take k xs) (l₂ := [y])).symm
      have h3 :
        ((List.take k xs) ++ [y]).sum
          = ((List.take k xs) ++ List.take 1 (List.drop k xs)).sum := by
        simpa [hcons]
      have h4 :
        ((List.take k xs) ++ List.take 1 (List.drop k xs)).sum
          = (List.take (k + 1) xs).sum := by
        have htakeadd :
            List.take (k + 1) xs
              = List.take k xs ++ List.take 1 (List.drop k xs) := by
          simpa using (List.take_add (l := xs) (i := k) (j := 1))
        simpa [htakeadd]
      calc
        cur + (rest).head!
            = cur + (List.drop i numbers).head! := by simpa [hrest_eq]
        _ = ((List.take k xs).sum) + (List.drop i numbers).head! := by
          simpa [hcur_sum, xs, k]
        _ = ((List.take k xs).sum) + y := h1
        _ = ((List.take k xs) ++ [y]).sum := h2
        _ = ((List.take k xs) ++ List.take 1 (List.drop k xs)).sum := h3
        _ = (List.take (k + 1) xs).sum := h4
        _ = (List.take (i + 1 - curStart) (List.drop curStart numbers)).sum := by
          simpa [xs, k, hkadd]
  )
  case hcur_suffix_max.loop_2 => (
  have hdne_drop_i : List.drop i numbers ≠ [] := by
    simpa [hrest_eq] using if_pos
  have hhead_eq : (rest).head! = (List.drop i numbers).head! := by
    simpa [hrest_eq]
  have hcases : start ≤ i ∨ start = i + 1 := by
    cases lt_or_ge i start with
    | inl hi_lt_start =>
      have h1 : i + 1 ≤ start := Nat.succ_le_of_lt hi_lt_start
      have h2 : start ≤ i + 1 := a_2
      exact Or.inr (le_antisymm h2 h1)
    | inr hge =>
      exact Or.inl hge
  cases hcases with
  | inl hstart_le_i =>
    have hdrop_eq :
        List.drop (i - start) (List.drop start numbers) = List.drop i numbers := by
      have hadd : start + (i - start) = i := by
        simpa [Nat.add_comm] using (Nat.sub_add_cancel hstart_le_i)
      simpa [List.drop_drop, hadd]
    have hlen : i + 1 - start = (i - start) + 1 := by
      simpa [Nat.succ_eq_add_one] using (Nat.succ_sub hstart_le_i)
    have hsplit :
        (List.take (i + 1 - start) (List.drop start numbers)).sum
          = (List.take (i - start) (List.drop start numbers)).sum + (List.drop i numbers).head! := by
      have hx : List.drop (i - start) (List.drop start numbers) ≠ [] := by
        exact fun h0 => hdne_drop_i (by simpa [hdrop_eq] using h0)
      have hsum := sum_take_succ_eq_sum_take_add_head! (l := List.drop start numbers) (n := i - start) (h := hx)
      simpa [hlen, hdrop_eq] using hsum
    have hpre_le : (List.take (i - start) (List.drop start numbers)).sum ≤ cur := by
      exact hcur_suffix_max start hstart_le_i
    calc
      (List.take (i + 1 - start) (List.drop start numbers)).sum
          = (List.take (i - start) (List.drop start numbers)).sum + (List.drop i numbers).head! := by
              simpa [hsplit]
      _ ≤ cur + (List.drop i numbers).head! := by
              exact add_le_add_right hpre_le _
      _ = cur + (rest).head! := by
              simpa [hhead_eq]
  | inr hstart_eq_succ =>
    have : i + 1 - start = 0 := by
      simpa [hstart_eq_succ, Nat.sub_self]
    have pos : 0 < cur + (rest).head! := by
      exact lt_of_not_ge if_neg
    have nonneg : 0 ≤ cur + (rest).head! := le_of_lt pos
    simpa [this] using nonneg
  )
  case hbest_sum.loop_2 => (
  have hdrop_ne' : List.drop i numbers ≠ [] := by
    simpa [hrest_eq] using if_pos
  have idx_eq : curStart + (i - curStart) = i := by
    simpa [Nat.add_comm] using Nat.sub_add_cancel hcurStart_le
  have hdrop_ne :
      List.drop (i - curStart) (List.drop curStart numbers) ≠ [] := by
    simpa [List.drop_drop, idx_eq] using hdrop_ne'
  have hsum :=
    sum_take_succ_of_drop_ne_nil (List.drop curStart numbers) (i - curStart) hdrop_ne
  have hlen_eq : i - curStart + 1 = i + 1 - curStart := by
    simpa [Nat.succ_eq_add_one] using (Nat.succ_sub hcurStart_le).symm
  simpa [hlen_eq, List.drop_drop, idx_eq, hcur_sum, hrest_eq] using hsum
  )
  case hprefix_max.loop_2 => (
  have hdrop_ne : List.drop i numbers ≠ [] := by
    simpa [hrest_eq] using if_pos
  have hpos_curx : 0 < cur + (List.drop i numbers).head! := by
    exact not_le.mp (by
      simpa [hrest_eq] using if_neg)
  by_cases hend_le_i : start + len ≤ i
  ·
    have hstart_le_i : start ≤ i := Nat.le_trans (Nat.le_add_right _ _) hend_le_i
    have hsum_le_max :
        (List.take len (List.drop start numbers)).sum ≤ max :=
      hprefix_max start len hstart_le_i hend_le_i
    have hmax_lt_curx_drop : max < cur + (List.drop i numbers).head! := by
      simpa [hrest_eq] using if_pos_1
    have hmax_le_curx_rest : max ≤ cur + rest.head! := by
      simpa [hrest_eq] using (le_of_lt hmax_lt_curx_drop)
    exact le_trans hsum_le_max hmax_le_curx_rest
  ·
    have hi_le_sum : i ≤ start + len := by
      cases Nat.le_total i (start + len) with
      | inl h => exact h
      | inr h =>
        exact (False.elim (hend_le_i h))
    have hi_lt_sum : i < start + len := Nat.lt_of_le_of_ne hi_le_sum (by
      intro h
      have hle : start + len ≤ i := by simpa [h] using le_of_eq (Eq.symm h)
      exact (hend_le_i hle).elim)
    have hge_succ : i + 1 ≤ start + len := Nat.succ_le_of_lt hi_lt_sum
    have hEq_end : start + len = i + 1 := Nat.le_antisymm a_3 hge_succ
    have hstart_le_or_eq := Nat.lt_or_eq_of_le a_2
    cases hstart_le_or_eq with
    | inr hstart_eq =>
      have hlen0 : len = 0 := by
        have := congrArg (fun t => t - start) hEq_end
        simpa [Nat.add_sub_cancel, hstart_eq] using this
      have : (List.take len (List.drop start numbers)).sum = 0 := by
        simpa [hlen0]
      have hnonneg_curx_rest : 0 ≤ cur + rest.head! := by
        simpa [hrest_eq] using (le_of_lt hpos_curx)
      simpa [this] using hnonneg_curx_rest
    | inl hstart_lt_succ =>
      have hstart_le_i : start ≤ i := Nat.lt_succ_iff.mp hstart_lt_succ
      have hlen_eq : len = i + 1 - start := by
        have := congrArg (fun t => t - start) hEq_end
        simpa [Nat.add_sub_cancel] using this
      have hS_le_cur :
          (List.take (i - start) (List.drop start numbers)).sum ≤ cur :=
        hcur_suffix_max start hstart_le_i
      have h_take_take :
          List.take (i - start)
            (List.take (i + 1 - start) (List.drop start numbers))
            = List.take (i - start) (List.drop start numbers) := by
        have hle : i - start ≤ i + 1 - start :=
          Nat.sub_le_sub_right (Nat.le_succ i) start
        simpa [List.take_take, Nat.min_eq_left hle]
      have h_drop_take :
          List.drop (i - start)
            (List.take (i + 1 - start) (List.drop start numbers))
            = List.take ((i + 1 - start) - (i - start))
                (List.drop (i - start) (List.drop start numbers)) := by
        simpa [List.drop_take]
      have hadd_eq : start + (i - start) = i := by
        have := Nat.sub_add_cancel hstart_le_i
        simpa [Nat.add_comm] using this
      have hdiff_one :
          (i + 1 - start) - (i - start) = 1 := by
        calc
          (i + 1 - start) - (i - start)
              = (i + 1) - (start + (i - start)) := by
                  simpa [Nat.sub_sub]
            _ = (i + 1) - i := by simpa [hadd_eq]
            _ = 1 := by simpa [Nat.add_comm] using Nat.add_sub_cancel i 1
      have hdropdrop :
          List.drop (i - start) (List.drop start numbers) = List.drop i numbers := by
        simpa [Nat.add_comm, hadd_eq] using (List.drop_drop (i - start) start numbers)
      have hdecomp :
          (List.take (i + 1 - start) (List.drop start numbers))
          = List.take (i - start)
              (List.take (i + 1 - start) (List.drop start numbers))
            ++
            List.drop (i - start)
              (List.take (i + 1 - start) (List.drop start numbers)) := by
        have := List.take_append_drop (i - start)
            (List.take (i + 1 - start) (List.drop start numbers))
        exact this.symm
      have hsum_decomp :
          (List.take (i + 1 - start) (List.drop start numbers)).sum
          =
          (List.take (i - start) (List.take (i + 1 - start) (List.drop start numbers))).sum
          +
          (List.drop (i - start) (List.take (i + 1 - start) (List.drop start numbers))).sum := by
        simpa [List.sum_append] using congrArg List.sum hdecomp
      have hsum_first :
          (List.take (i - start) (List.take (i + 1 - start) (List.drop start numbers))).sum
          = (List.take (i - start) (List.drop start numbers)).sum := by
        simpa [List.take_take, Nat.min_eq_left (Nat.sub_le_sub_right (Nat.le_succ i) start)]
          using congrArg List.sum h_take_take
      have hsum_second :
          (List.drop (i - start) (List.take (i + 1 - start) (List.drop start numbers))).sum
          = (List.take 1 (List.drop i numbers)).sum := by
        have := congrArg List.sum h_drop_take
        simpa [hdiff_one, hdropdrop] using this
      have hsum_take1_head :
          (List.take 1 (List.drop i numbers)).sum = (List.drop i numbers).head! := by
        cases hdi : List.drop i numbers with
        | nil =>
          cases hdrop_ne (by simpa [hdi])
        | cons y ys =>
          simp [hdi]
      have hsum_eq :
          (List.take (i + 1 - start) (List.drop start numbers)).sum
          =
          (List.take (i - start) (List.drop start numbers)).sum
          + (List.drop i numbers).head! := by
        calc
          (List.take (i + 1 - start) (List.drop start numbers)).sum
              =
              (List.take (i - start) (List.take (i + 1 - start) (List.drop start numbers))).sum
              + (List.drop (i - start) (List.take (i + 1 - start) (List.drop start numbers))).sum := by
                exact hsum_decomp
          _ = (List.take (i - start) (List.drop start numbers)).sum
              + (List.take 1 (List.drop i numbers)).sum := by
                simpa [hsum_first, hsum_second]
          _ = (List.take (i - start) (List.drop start numbers)).sum
              + (List.drop i numbers).head! := by
                simpa [hsum_take1_head]
      have hS_le_cur' :
          (List.take (i - start) (List.drop start numbers)).sum ≤ cur := hS_le_cur
      have : (List.take (i + 1 - start) (List.drop start numbers)).sum ≤ cur + (List.drop i numbers).head! := by
        have hadded := add_le_add_right hS_le_cur' ((List.drop i numbers).head!)
        simpa [hsum_eq] using hadded
      simpa [hlen_eq, hrest_eq] using this
  )
  case decreasing.loop_2 => (
  cases rest with
  | nil =>
    cases if_pos rfl
  | cons x xs =>
    have h :
      (WithName.mk' xs (Lean.Name.anonymous.mkStr "rest")).erase.length = xs.length := by
      rfl
    simpa [List.length, h] using Nat.lt_succ_self xs.length
  )
  case hi_len.loop_3 => (
  have hpos : 0 < rest.length := List.length_pos_of_ne_nil if_pos
  have hge1 : 1 ≤ rest.length := Nat.succ_le_of_lt hpos
  have hlen_tail :
      (WithName.mk' rest.tail (Lean.Name.anonymous.mkStr "rest")).erase.length
        = rest.length - 1 := by
    have heq : (WithName.mk' rest.tail (Lean.Name.anonymous.mkStr "rest")).erase.length
                 = rest.tail.length := rfl
    simpa [heq, List.length_tail]
  have hsub : rest.length - 1 + 1 = rest.length := Nat.sub_add_cancel hge1
  calc
    i + 1 + (WithName.mk' (rest).tail (Lean.Name.anonymous.mkStr "rest")).erase.length
        = i + 1 + (rest.length - 1) := by simpa [hlen_tail]
    _   = i + ((rest.length - 1) + 1) := by
            simpa [Nat.add_assoc] using
              (Nat.add_right_comm i 1 (rest.length - 1))
    _   = i + rest.length := by
            simpa [hsub]
    _   = numbers.length := by
            simpa [hi_len]
  )
  case hrest_eq.loop_3 => (
  try simp_all ; try grind
  )
  case hi_bounds.loop_7 => (
  have hpos : 0 < rest.length := List.length_pos_of_ne_nil if_pos
  have hone : 1 ≤ rest.length := Nat.succ_le_of_lt hpos
  have hle : i + 1 ≤ i + rest.length := Nat.add_le_add_left hone i
  simpa [hi_len] using hle
  )
  case hcur_sum.loop_3 => (
  have hrest_ne : List.drop i numbers ≠ [] := by
    simpa [hrest_eq] using if_pos
  have hadd : curStart + (i - curStart) = i := Nat.add_sub_of_le hcurStart_le
  have hdropdrop :
      List.drop (i - curStart) (List.drop curStart numbers)
        = List.drop i numbers := by
    simpa [hadd, Nat.add_comm, Nat.add_left_comm, Nat.add_assoc] using
      (List.drop_drop (l := numbers) (n := curStart) (m := i - curStart)).symm
  have hdropne2 :
      List.drop (i - curStart) (List.drop curStart numbers) ≠ [] := by
    simpa [hdropdrop] using hrest_ne
  have hsucc : i + 1 - curStart = (i - curStart) + 1 := by
    simpa [Nat.succ_eq_add_one] using (Nat.succ_sub hcurStart_le)
  calc
    cur + (rest).head!
        = ((List.take (i - curStart) (List.drop curStart numbers))).sum
          + ((List.drop i numbers).head!) := by
          simpa [hcur_sum, hrest_eq]
    _ = ((List.take (i - curStart) (List.drop curStart numbers))).sum
          + ((List.drop (i - curStart) (List.drop curStart numbers)).head!) := by
          simpa [hdropdrop]
    _ = ((List.take ((i - curStart) + 1) (List.drop curStart numbers))).sum := by
          have hsum :=
            sum_take_succ_eq_sum_take_add_head!
              (l := List.drop curStart numbers) (n := i - curStart) (h := hdropne2)
          simpa using hsum.symm
    _ = (List.take (i + 1 - curStart) (List.drop curStart numbers)).sum := by
          simpa [hsucc]
  )
  case hcur_suffix_max.loop_3 => (
  have hrestNE : List.drop i numbers ≠ [] := by simpa [hrest_eq] using if_pos
  cases hdi : List.drop i numbers with
  | nil =>
    cases hrestNE (by simpa [hdi])
  | cons x xs =>
    have hhead_drop : (List.drop i numbers).head! = x := by simp [hdi]
    have hhead_rest : rest.head! = x := by simpa [hrest_eq, hdi] using hhead_drop
    have hcase : start ≤ i ∨ start = i + 1 := by
      cases lt_or_eq_of_le a_2 with
      | inl hlt =>
        exact Or.inl (Nat.lt_succ_iff.mp hlt)
      | inr heq =>
        exact Or.inr heq
    cases hcase with
    | inl hsi =>
      set L := List.drop start numbers
      set n1 := i - start
      have hLlen : L.length = numbers.length - start := by
        simpa [L] using List.length_drop start numbers
      have hn1_le_L : n1 ≤ L.length := by
        have : i - start ≤ numbers.length - start := Nat.sub_le_sub_right a_1 _
        simpa [hLlen, n1] using this
      have hlen_take : (List.take n1 L).length = n1 := by
        simpa [List.length_take, Nat.min_eq_left hn1_le_L]
      have hsub_eq : i + 1 - start = n1 + 1 := by
        have : (i - start) + start = i := Nat.sub_add_cancel hsi
        have : i + 1 = start + (n1 + 1) := by
          have : i + 1 = ((i - start) + start) + 1 := by simpa [this]
          simpa [n1, Nat.add_comm, Nat.add_left_comm, Nat.add_assoc] using this
        simpa [this, Nat.add_sub_cancel, n1]
      have hdropdrop : List.drop n1 L = List.drop i numbers := by
        have : List.drop n1 (List.drop start numbers) = List.drop (n1 + start) numbers := by
          simpa [Nat.add_comm] using (List.drop_drop numbers start n1)
        simpa [L, n1, Nat.sub_add_cancel hsi] using this
      have ht_decomp :
        List.take (n1 + 1) L
          = List.take n1 L ++ List.take ((n1 + 1) - (List.take n1 L).length) (List.drop n1 L) := by
        have h := List.take_append (l₁ := List.take n1 L) (l₂ := List.drop n1 L) (i := n1 + 1)
        simpa [List.take_append_drop, List.take_take, Nat.min_eq_right (Nat.le_succ n1)] using h
      have hsum_eq :
        (List.take (i + 1 - start) (List.drop start numbers)).sum
          = (List.take (i - start) (List.drop start numbers)).sum
            + (List.take 1 (List.drop i numbers)).sum := by
        calc
          (List.take (i + 1 - start) (List.drop start numbers)).sum
              = (List.take (n1 + 1) L).sum := by
                simp [L, n1, hsub_eq]
          _ = (List.take n1 L ++ List.take ((n1 + 1) - (List.take n1 L).length) (List.drop n1 L)).sum := by
                simpa [L, n1] using (congrArg List.sum ht_decomp)
          _ = (List.take n1 L).sum + (List.take ((n1 + 1) - (List.take n1 L).length) (List.drop n1 L)).sum := by
                simpa using (List.sum_append (List.take n1 L) (List.take ((n1 + 1) - (List.take n1 L).length) (List.drop n1 L)))
          _ = (List.take n1 L).sum + (List.take 1 (List.drop n1 L)).sum := by
                simp [hlen_take, Nat.add_comm, Nat.add_left_comm, Nat.add_assoc]
          _ = (List.take (i - start) (List.drop start numbers)).sum
              + (List.take 1 (List.drop i numbers)).sum := by
                simp [L, n1, hdropdrop]
      have hpref_le : (List.take (i - start) (List.drop start numbers)).sum ≤ cur :=
        hcur_suffix_max start hsi
      calc
        (List.take (i + 1 - start) (List.drop start numbers)).sum
            = (List.take (i - start) (List.drop start numbers)).sum
              + (List.take 1 (List.drop i numbers)).sum := hsum_eq
        _ = (List.take (i - start) (List.drop start numbers)).sum + x := by
              simp [hdi]
        _ ≤ cur + x := by
              exact add_le_add_right hpref_le x
        _ = cur + (rest).head! := by
              simpa [hhead_rest]
    | inr heq =>
      have hsum0 : (List.take (i + 1 - start) (List.drop start numbers)).sum = 0 := by
        simp [heq]
      have hpos : 0 < cur + (rest).head! := by
        exact lt_of_not_ge if_neg
      have hle : 0 ≤ cur + (rest).head! := le_of_lt hpos
      simpa [hsum0]
  )
  case hprefix_max.loop_3 => (
  classical
  -- From the loop condition, rest = numbers.drop i is nonempty
  have hrest_nonempty : List.drop i numbers ≠ [] := by
    simpa [hrest_eq] using if_pos
  -- Branch on whether start = i+1
  by_cases hstart_eq_i1 : start = i + 1
  · subst hstart_eq_i1
    have hle : i + 1 + len ≤ i + 1 := by simpa using a_3
    have hlen_le_zero : len ≤ 0 := by simpa using hle
    have hlen_zero : len = 0 := Nat.le_zero.mp hlen_le_zero
    simp [hlen_zero, hmax_nonneg]
  ·
    have hstart_le_i : start ≤ i := by
      have hlt : start < i + 1 := Nat.lt_of_le_of_ne a_2 hstart_eq_i1
      simpa [Nat.lt_succ_iff] using hlt
    by_cases h_le_i : start + len ≤ i
    · exact hprefix_max start len hstart_le_i h_le_i
    · -- so start + len = i + 1
      have hi1_le : i + 1 ≤ start + len := Nat.succ_le_of_lt (Nat.lt_of_not_ge h_le_i)
      have h_eq : start + len = i + 1 := Nat.le_antisymm a_3 hi1_le
      -- set k = i - start
      let k := i - start
      have hk : k + start = i := Nat.sub_add_cancel hstart_le_i
      -- len = k + 1
      have hlen_eq2 : len = k + 1 := by
        have h1 : start + len = start + (k + 1) := by
          have htmp : i = start + k := by
            have : i = k + start := by simpa using hk.symm
            simpa [Nat.add_comm] using this
          simpa [htmp, Nat.add_assoc] using h_eq
        exact Nat.add_left_cancel h1
      -- show drop k (drop start numbers) is nonempty
      have dd : List.drop k (List.drop start numbers) = List.drop (start + k) numbers := by
        simpa [Nat.add_comm, Nat.add_left_comm, Nat.add_assoc] using List.drop_drop k start numbers
      have hdrop_nonempty : List.drop k (List.drop start numbers) ≠ [] := by
        have hk_comm : start + k = i := by simpa [Nat.add_comm] using hk
        have : List.drop (start + k) numbers ≠ [] := by simpa [hk_comm] using hrest_nonempty
        simpa [dd] using this
      -- cur + head! ≤ max (since this branch didn't update max)
      have hcurhead_le_max0 :
          cur + rest.head! ≤ max := not_lt.mp if_neg_1
      have hcurhead_le_max :
          cur + (List.drop i numbers).head! ≤ max := by
        simpa [hrest_eq] using hcurhead_le_max0
      -- Bind old part by cur
      have hprefix_end_le_cur :
          (List.take k (List.drop start numbers)).sum ≤ cur := by
        simpa [k] using hcur_suffix_max start hstart_le_i
      -- Combine using the generic take-succ sum lemma
      have : (List.take len (List.drop start numbers)).sum ≤ max := by
        calc
          (List.take len (List.drop start numbers)).sum
              = (List.take (k + 1) (List.drop start numbers)).sum := by
                simpa [hlen_eq2]
          _ = (List.take k (List.drop start numbers)).sum +
              (List.drop k (List.drop start numbers)).head! := by
                simpa using (sum_take_succ_eq_sum_take_add_head!
                              (l := List.drop start numbers) (n := k) (h := hdrop_nonempty))
          _ ≤ cur + (List.drop k (List.drop start numbers)).head! :=
                add_le_add_right hprefix_end_le_cur _
          _ ≤ max := by
                -- rewrite the head via dd and hk
                have hk_comm : start + k = i := by simpa [Nat.add_comm] using hk
                have : (List.drop k (List.drop start numbers)).head!
                        = (List.drop (start + k) numbers).head! := by simpa [dd]
                have : cur + (List.drop (start + k) numbers).head! ≤ max := by
                  simpa [hk_comm] using hcurhead_le_max
                simpa [this]
      exact this
  )
  case decreasing.loop_3 => (
  cases rest with
  | nil =>
    cases if_pos rfl
  | cons x xs =>
    have h : (WithName.mk' xs (Lean.Name.anonymous.mkStr "rest")).erase = xs := rfl
    simpa [h] using Nat.lt_succ_self (List.length xs)
  )
  case hcur_suffix_max.entry => (
  try simp_all ; try grind
  )
  case hprefix_max.entry => (
  try simp_all ; try grind
  )
  case «ensures» => (
  rcases (by simpa using i_7) with ⟨hbLenEq, hbStartEq, hcurEq, hcurStartEq, hiEqI, hmaxEq, hrest1Eq⟩
  have hiEq : i = numbers.length := by
    simpa [hdone] using hi_len
  have hStartLE : bestStart ≤ numbers.length := by simpa [hiEq] using hbest_start_in_prefix
  have hEndLE' : bestStart + bestLen ≤ numbers.length := by simpa [hiEq] using hbest_end_in_prefix
  simp [hmaxEq, hbLenEq, hbStartEq, hiEq]
  refine ⟨bestStart, Nat.lt_succ_of_le hStartLE, ⟨bestLen, ?_, ?_⟩⟩
  have hRHSrewrite : bestStart + (numbers.length - bestStart) = numbers.length := by
    simpa [Nat.add_comm] using Nat.sub_add_cancel hStartLE
  have hineq : bestStart + bestLen ≤ bestStart + (numbers.length - bestStart) := by
    simpa [hRHSrewrite] using hEndLE'
  have hLenLE : bestLen ≤ numbers.length - bestStart :=
    Nat.le_of_add_le_add_left hineq
  exact Nat.lt_succ_of_le hLenLE
  simpa [hmaxEq] using hbest_sum
  )
  case ensures_1 => (
  have hi_eq : i = numbers.length := by
    simpa [hdone] using hi_len
  have hmax_eq : max = i_6 := by
    have := congrArg (fun (p : _) => p.snd.snd.snd.snd.snd.fst) i_7
    simpa using this
  refine (List.all_eq_true.mpr ?H)
  intro x hx
  rcases List.mem_flatMap.1 hx with ⟨start, hstart_mem, hx_mem⟩
  rcases List.mem_map.1 hx_mem with ⟨len, hlen_mem, rfl⟩
  have hstart_lt : start < numbers.length + 1 := List.mem_range.1 hstart_mem
  have hlen_lt   : len < numbers.length - start + 1 := List.mem_range.1 hlen_mem
  have hstart_le : start ≤ numbers.length := Nat.lt_succ_iff.mp hstart_lt
  have hlen_le   : len ≤ numbers.length - start := Nat.lt_succ_iff.mp hlen_lt
  have hstart_len_le : start + len ≤ numbers.length := by
    have h := Nat.add_le_add_left hlen_le start
    have hsum : start + (numbers.length - start) = numbers.length := Nat.add_sub_of_le hstart_le
    simpa [hsum] using h
  have hsum_le_max : (List.take len (List.drop start numbers)).sum ≤ max := by
    have h1 : start ≤ i := by simpa [hi_eq] using hstart_le
    have h2 : start + len ≤ i := by simpa [hi_eq] using hstart_len_le
    exact hprefix_max start len h1 h2
  have hsum_le_i6 : (List.take len (List.drop start numbers)).sum ≤ i_6 := by
    simpa [hmax_eq] using hsum_le_max
  exact (decide_eq_true_iff).2 hsum_le_i6
  )
\end{lstlisting}

\section{Example proof: auto-active style}
The following proof found by GPT-5 in a sequential agent scaffolding demonstrates the \emph{auto-active} proving style where the user provides invariants, lemmas and solver hints and then hands over the remainder of the proof to SMT solvers (CVC5 via \texttt{loom\_solve} and Lean's automation tools (\texttt{simp\_all}).

\lstinputlisting[caption={Implementation and correctness proof for a power of two check algorithm (\texttt{verina\_advanced\_23})},label={lst:pow2}, captionpos=t, language=lean, breaklines=true, style=lean]{listings/pow2.lean}

\subsection{Termination in the Clever benchmark}
In the Clever benchmark, termination is formulated using a ``the output exists'' formulation such as in the following example.
\begin{lstlisting}[language=lean, breaklines=true, style=lean]
def problem_spec
  (implementation: List Int → Int → Bool)
  (q: List Int) (w: Int) :=
  let spec (result : Bool) :=
    result ↔ (List.Palindrome q) ∧ (List.sum q ≤ w)
  ∃ result, implementation q w = result ∧ spec result
\end{lstlisting}
We note here that this is unnecessary because functions such as \texttt{implementation: List Int → Int → Bool} in Lean are always total (proved to terminate by the equation compiler), as non-terminating functions would lead to inconsistencies in the logic (example: a non-terminating function which produces proofs of \texttt{False}). The existential statement \texttt{$\exists$ result, f x = result} can always trivially be proved with \texttt{ ⟨f x, rfl⟩}.

On the other hand, for imperative code, proof of termination is typically more involved, and covered by \texttt{decreasing} annotations in Velvet's syntax and the corresponding proof obligations.

\section{Prompt}
\label{app:prompts}

We layout the procedure of the evolution of the prompts used in our framework. We provide Gemini 2.5 Pro with the Velvet documentation\footnote{\url{https://github.com/verse-lab/loom/blob/master/CaseStudies/Velvet/velvet_documentation.md}} and instruct it to write a prompt for an agentic system which do not have prior knowledge of it. We use this version of prompt for experiments during the development of the framework. The prompt is refined by us through manually looking at early failure cases. This version of prompt gives 53.44\% of pass@4 on Verina with GPT-5 sequential agent.

In later iterations, we provide Claude Code with the whole trajectories directory and instruct it to further enhance the prompt. Claude Code is able to dynamically explore trajectory files, write scripts to gather statistics and spot several failing patterns, such as the usage of sorry and non-existing lemma, and add extra instruction, e.g., retry different strategies after consistent failures. With the enhanced prompt, pass@4 on Verina with GPT-5 sequntial agent increases to 56.08\%.

However, this step also introduces some inconsistencies and misconceptions about Loom/Lean. We conclude the final step by looking at the prompt manually and cleaning up these inconsistencies. The final version of the prompt, shown in Listing~\ref{lst:prompt_linear}, results in 58.73\% pass@4 on Verina with the same setup.

\lstinputlisting[caption={Final prompt for sequential agent, with highlights showing the diff compared with the first version},label={lst:prompt_linear}, captionpos=t, language=lean, breaklines=true, style=lean]{listings/prompt_linear.md}

\end{document}